\documentclass[fleqn,usenatbib]{mnras}
\usepackage{newtxtext,newtxmath}

\usepackage[T1]{fontenc}
\usepackage{ae,aecompl}

\usepackage{graphicx}	
\usepackage{amsmath}	
\usepackage{amssymb}	
\usepackage{lastpage}

\newcommand{\m}{\checkmark}
\graphicspath{{./}{figures/}}

\title[Dwarf Galaxy Kinematics]{AGN and Dwarf Galaxy Gas Kinematics}
\author[0000-0002-5253-9433]{
Christina M. Manzano-King,$^{1}$\thanks{E-mail: cking012@ucr.edu (CMK)}
Gabriela Canalizo$^{1}$
\newauthor
\\
$^{1}$University of California, Riverside, CA, USA\\
}

\date{Accepted 2020 August 25. Received 2020 August 23; in original form 2020 July 24}

\pubyear{2020}

\begin{document}
\label{firstpage}
\maketitle

\begin{abstract}
We present spatially resolved kinematic measurements of stellar and ionized gas components of dwarf galaxies in the stellar mass range $10^{8.5} - 10^{10} M_{\odot}$, selected from SDSS DR7 and DR8 and followed-up with Keck/LRIS spectroscopy.  We study the potential effects of active galactic nuclei (AGN) on galaxy-wide gas kinematics by comparing rotation curves of 26 galaxies containing AGN, and 19 control galaxies with no optical or infrared signs of AGN.  We find a strong association between AGN activity and disturbed gas kinematics in the host galaxies. While star forming galaxies in this sample tend to have orderly gas discs that co-rotate with the stars, 73\% of the AGN have disturbed gas. We find five out of 45 galaxies have gaseous components in counter-rotation with their stars, and all galaxies exhibiting counter-rotation contain AGN.  Six out of seven isolated galaxies with disturbed ionized gas host AGN.  At least three AGN fall clearly below the stellar-halo mass relation, which could be interpreted as evidence for ongoing star formation suppression.  Taken together, these results provide new evidence supporting the ability of AGN to influence gas kinematics and suppress star formation in dwarf galaxies.  This further demonstrates the importance of including AGN as a feedback mechanism in galaxy formation models in the low-mass regime.
\end{abstract}





\section{Introduction}

In the $\Lambda$CDM model of structure formation, dark matter halos form from the gravitational collapse of primordial density fluctuations. Within these halos, baryons collapse into a rotating disc with the same angular momentum as the dark matter halo.  With sufficient radiative cooling, the gas is able to collapse and form stars, resulting in a co-rotating disc of gas and stars within a dark matter halo with a density profile which can be approximated by a simple formula (Navarro, Frenk \& White 1996, 1997).  On large scales, the $\Lambda$CDM model of structure formation agrees well with observations of massive galaxies and clusters.  

However, large discrepancies in the low-mass regime raise doubts about $\Lambda$CDM.  For example, there is a large disagreement between the number of small dark matter halos and observed dwarfs \citep[missing satellite problem,][]{Moore1999,Klypin1999}.  There is also a conspicuous absence of observed large satellites compared to predictions \citep[too big to fail problem,][]{Read2006,Boylan-Kolchin2011}.  Finally, rotation profiles of dwarfs show a variety of inner slopes, indicating a diversity of dark matter halo profiles, many of which are in disagreement with the predicted NFW dark matter density profile \citep[cusp vs. core problem,][]{Flores1994,Simon2005,Oh2011}.  These departures from the $\Lambda$CDM model have prompted some to look into warm \citep[e.g.][]{Lovell2014} or self-interacting dark matter \citep[e.g.][] {Rocha2013}.  A promising alternative to re-thinking the nature of dark matter is to investigate the effects of baryonic feedback on star formation.  
        
High resolution simulations conclude that baryonic processes can be used to reconcile the observed properties of dwarf galaxies with $\Lambda$CDM \citep{Wetzel2016ApJ...827L..23W}.  Theoretical work commonly attributes star formation suppression in dwarf galaxies to heating from the UV background during reionization  \citep[e.g.][]{Katz2020}, stellar radiation \citep[e.g.][]{Emerick2018}, and supernovae feedback \citep[e.g.][]{Hu2019}.  Indeed, powerful star-formation driven outflows are observed in starbursting dwarf galaxies \citep[][]{Martin1998,Strickland_Stevens_2000}, and some observational evidence suggests that stellar feedback may dominate in dwarf galaxies \citet{Navarro2018}.

Stellar feedback is only part of the picture, as the role of AGN feedback in dwarfs becomes harder to ignore.  Evidence of AGN via optical and infrared (IR) indicators has been detected in hundreds of nearby dwarf galaxies \citep[][]{Reines2013,Moran2014,Sartori2015,Baldassare2020,Birchall2020}.  \citet{Kaviraj2019} report the IR-selected AGN occupation fraction in high mass galaxies to be $1-3\%$, while the same criteria yield a $10-30\%$ fraction in dwarf galaxies ($M_* \sim 10^{8-10} M_\odot$).  Given that there are several factors that hinder the detection of AGN in dwarfs \citep{Satyapal2018, Cann2019}, this large AGN fraction can be regarded as a lower limit.  These studies suggest that AGN are common and potentially important phenomena in the low mass regime.  

Additionally, observational evidence of AGN-driven outflows in dwarf galaxies has begun to surface \citep{Penny2018,Bradford2018,Dickey2019,Manzano2019}.  Each of these studies present observations of AGN coexisting with kinematically disturbed gas.  Such disturbances are distinguished by broadened components in their emission line profiles and velocity measurements indicating gas that is disconnected or even counter-rotating with respect to their stars. For example, \citet{Starkenburg2019} (henceforth S19) use data from Illustris \citep{Vogelsberger2014a,Genel2014} to investigate the origin of star-gas counter-rotation in dwarf galaxies.  By examining the evolutionary history of simulated galaxies with counter-rotating gas, they found that such counter-rotating components require removal of the original gas reservoir and re-accretion of new gas, with misaligned angular momentum.  S19 identify two plausible mechanisms for gas removal: stripping from an encounter with a neighbouring galaxy, or AGN outflow events.

The role of AGN feedback in galaxy formation and evolution is not well understood in general.  The shallow potential wells of dwarf galaxies leave them particularly susceptible to feedback, making them ideal laboratories to study how the energy output of an AGN couples to the gas in the host galaxy.  The discovery of AGN-driven outflows in the dwarf regime challenges current conceptions of feedback in dwarf galaxies and raises the question of whether the gas in these outflows permanently escapes the halo.  In this paper, we explore the connection between AGN and kinematically disturbed gas and present further evidence of AGN feedback in dwarf galaxies.

Throughout the paper we assume the cosmological model $H_0 = 71\,\rm{km}\,\rm{s}^{-1}\,\rm{Mpc}^{-1}$, $\Omega_m = 0.27$, and $\Omega_\Lambda = 0.73$.

\section{Data}\label{sec:Data}

\begin{figure}
\includegraphics[width=0.49\textwidth]{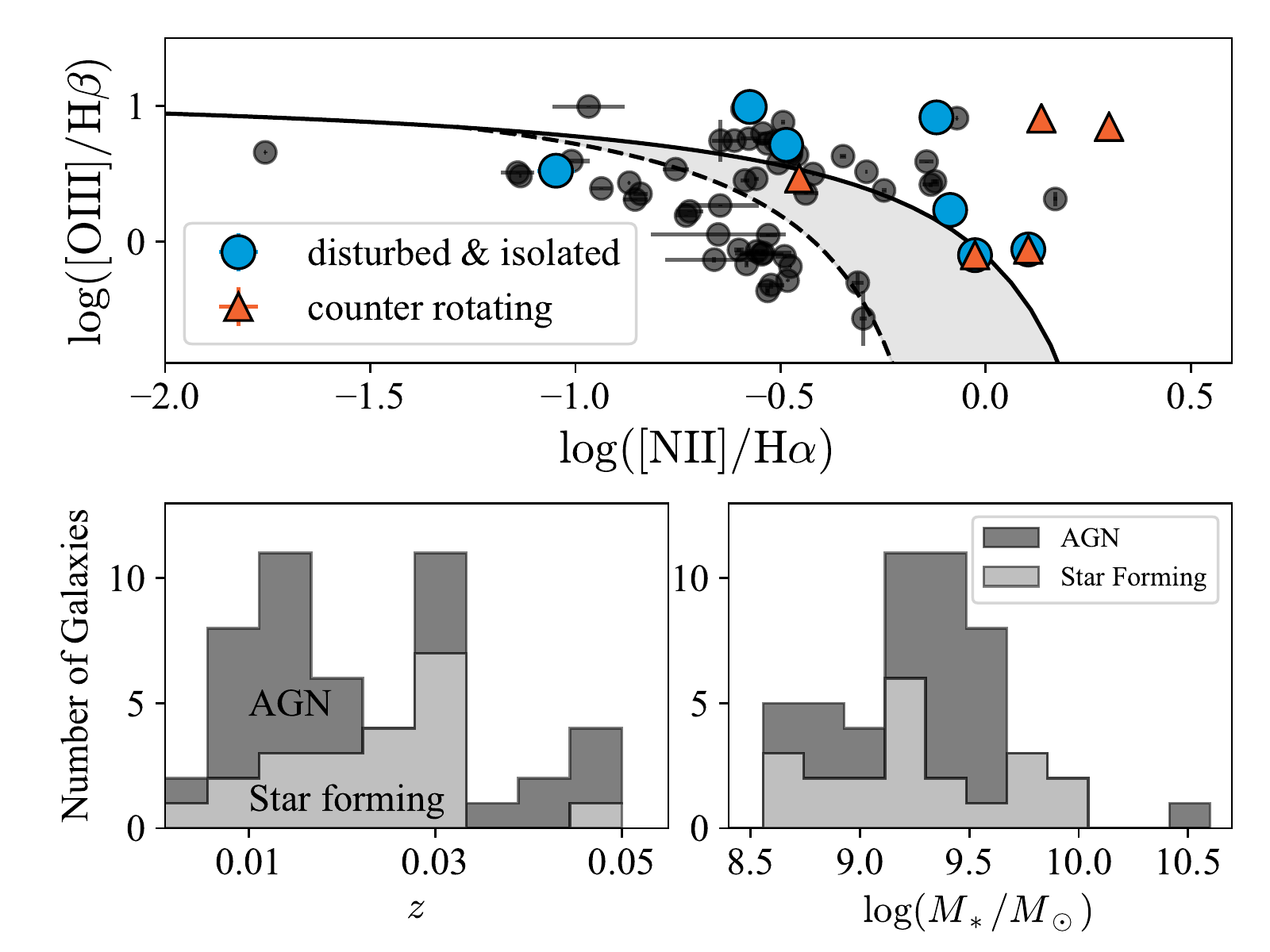}
\caption{Top: BPT diagnostic for all 50 galaxies in this sample. The dotted line denotes the \citet{kauffmann2003} classification cutoff, and the solid line is the \citet{kewley2001} maximum starburst line. Blue circles indicate isolated galaxies with disturbed gas and with no neighbours of comparable mass (i.e. the neighbouring galaxy is at least 0.75 times the stellar mass of the dwarf) within 1.5 Mpc (see Section \ref{sec:environment}).  Galaxies with counter-rotating stellar and gas components are marked as orange triangles.
Bottom:Stacked histograms of redshift and stellar mass of the galaxies presented in this paper. Dark gray represents AGN hosts and light gray represents star forming galaxies.
}
\label{fig:bpthist}
\end{figure}

\begin{center}
\begin{figure}
\includegraphics[width=\linewidth,clip]{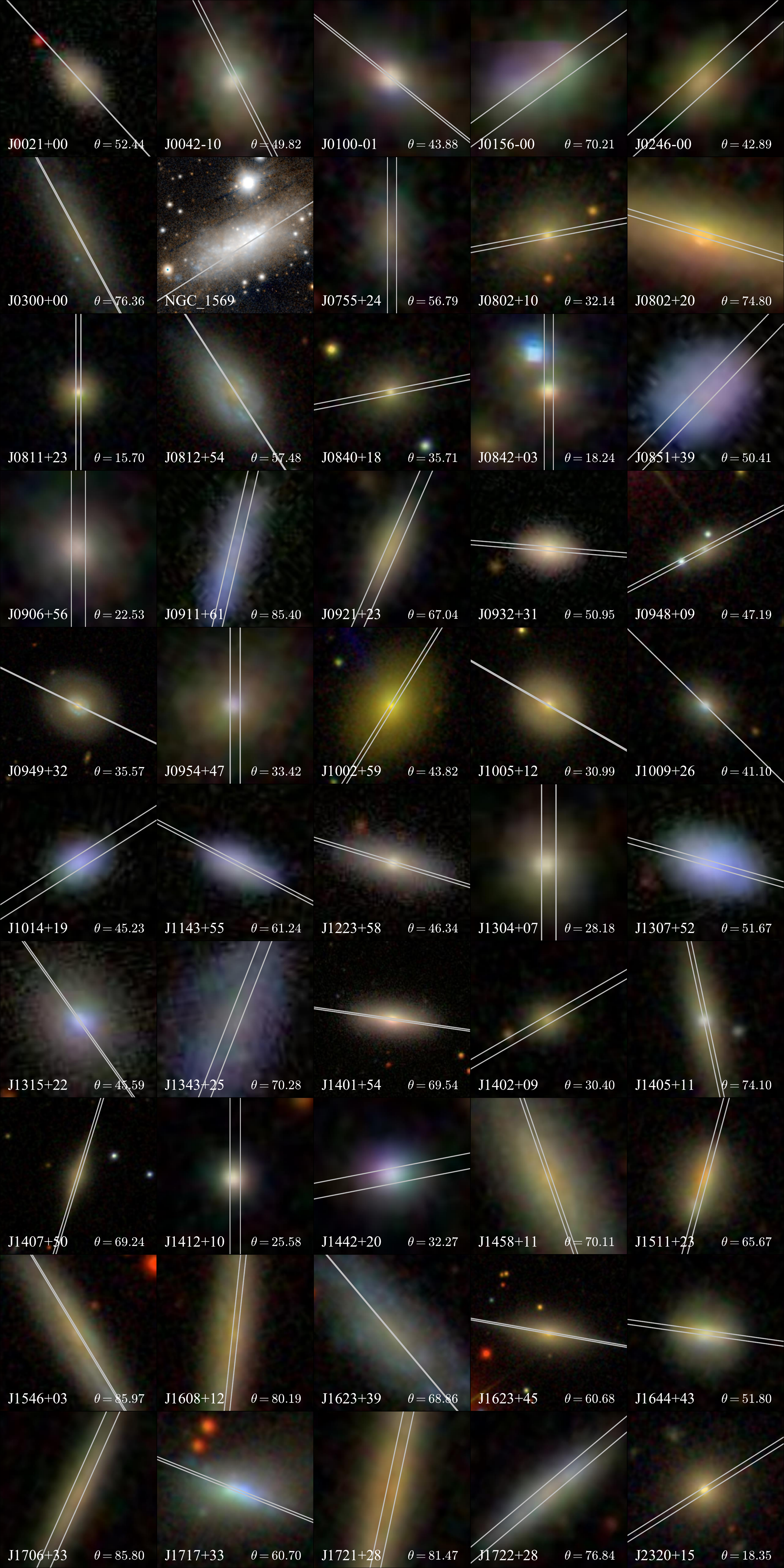}
\caption{colour images of all 50 dwarf galaxies in our sample.  All images were generated using the SDSS DR12 finding chart tool, with the exception of NGC 1569, which is outside of the SDSS footprint.  The NGC 1569 thumbnail (second row, second column) is a PanSTARS z,g band colour image rendered in the Aladin Lite Viewer with a 3' field of view ($\sim 1\,$kpc on a side).  Each SDSS image is scaled to 10 kpc on a side and the placement of the 1 arcsecond-wide slit is shown in light gray.  
}
\label{fig:thumbs}
\end{figure}
\end{center}

Our complete sample contains 50 nearby ($z=0.05$) dwarf galaxies (roughly $M_\star < 10^{10} M_\odot$) drawn from the Sloan Digital Sky Survey (SDSS).  SDSS is 95\% complete to $r$ band magnitude of 22.2, which is well below the magnitude range covered in this sample ($r\approx 12-17)$.  Twenty-nine of the galaxies in our sample are classified as AGN based on emission line flux ratios that fall above the \citet{kauffmann2003} star forming sequence on the Baldwin, Phillips \& Terlevich (BPT) diagnostic \citep{baldwin1981} or the presence of detectable \ion{He}{II} emission \citep{shirazi2012}.  This sample of 29 dwarf galaxies hosting AGN was selected from the parent samples of \citet{Reines2013,Moran2014,Sartori2015,Oh2015} based on right ascensions that could be observed during our allotted Keck time.   In order to facilitate spatially resolved kinematic measurements, we prioritized spatially extended galaxies and excluded face-on galaxies whenever possible.  

To enable comparison between galaxies with and without AGN, we selected a control sample of 21 star forming galaxies from SDSS Data Release 8, based on their absence of optical and infrared (IR) signatures of AGN.  The control sample was selected by applying the same stellar mass and redshift cuts used to build the AGN sample, then excluding objects with emission line flux ratios falling above the star forming sequence on the BPT diagram.  We also excluded all potential AGN using the WISE mid-IR colour criteria \citep{Jarrett2011, stern2012} and further excluded all galaxies with detectable \ion{He}{II} emission.  The BPT diagnostic diagram and distributions of the redshifts and stellar masses of our full sample are shown in Figure \ref{fig:bpthist}.

We collected long-slit spectroscopy of 50 galaxies using the Low Resolution Imaging Spectrometer on the Keck I telescope \citep[LRIS][]{Oke_1995,Rockosi2010}. 
Observation dates, conditions, and spectrograph setups for each object are listed in Table \ref{tab:obs_setups}.  The DIMM seeing (Full Width at Half Maximum value of a star observed at zenith, at 5000\AA) is listed for each night. The slit position angles (PA) were determined by fitting $r-$band SDSS photometry using the IRAF ellipse task.  By placing the slit along the semi-major axes determined by these ellipse fits, as shown in Figure \ref{fig:thumbs}, we obtained spatially resolved spectra, presumably perpendicular to the rotation axis of each galaxy.

\newpage
\begin{table*}
\centering
\begin{tabular}{| l c c c c c c c c c c c|}
\hline
Name in SDSS & Observation Date  & Seeing  & slit PA& red setup & exposure time \\ 
& & $\arcsec$ & CCW N&&  s \\ 
\hline
J002145.80+003327.3   & 2015-12-05 &  0.59  &  42.4   & B & 1200 \\ 
J004214.99$-$104415.0 & 2015-12-04 &  0.55  &  26.9   & B & 1200 \\ 
J010005.94$-$011059.0 & 2015-12-04 &  0.55  &  51.2   & B & 1200 \\ 
J015645.30-003737.8   & 2015-12-04 &  0.55  &  126    & B & 1200 \\ 
J024656.39-003304.8*  & 2015-12-04 &  0.55  &  131.9  & B & 1200 \\ 
J030040.20+000113.3   & 2015-12-04 &  0.55  &  28.3   & B & 2400 \\ 
NGC 1569*             & 2015-12-04 &  0.55  &  122.5  & B & 2400 \\ 
J075538.19+240103.5   & 2015-12-04 &  0.55  &  0      & B & 2400 \\ 
J080212.06+103234.1   & 2015-12-05 &  0.59  &  100.7  & B & 2400 \\ 
J080228.83+203050.2   & 2015-12-04 &  0.55  &  73.3   & B & 2400 \\ 
J081145.29+232825.7   & 2015-12-04 &  0.55  &  0      & B & 2400 \\ 
J081256.37+545808.4   & 2015-12-05 &  0.59  &  32.77  & B & 1200 \\ 
J084025.54+181858.9   & 2015-12-04 &  0.55  &  101    & B & 2400 \\ 
J084234.51+031930.7   & 2015-12-05 &  0.59  &  0      & B & 1200 \\ 
J085125.81+393541.7   & 2015-03-24 &  0.70  & -43.96  & C & 1200 \\ 
J090613.75+561015.5   & 2015-12-04 &  0.55  &  0      & B & 2400 \\ 
J091122.24+615245.2   & 2015-03-24 &  0.70  &  -13    & C & 2400 \\ 
J092149.44+233438.7   & 2015-12-05 &  0.59  &  156.3  & B & 1200 \\ 
J093251.11+314145.0   & 2015-03-25 &  0.58  &  -94.6  & B & 1200 \\ 
J094800.79+095815.4   & 2015-12-05 &  0.59  &  117.8  & B & 1200 \\ 
J094941.20+321315.9   & 2015-12-04 &  0.55  &  64     & B & 1200 \\ 
J095418.16+471725.1   & 2015-12-05 &  0.59  &  0      & B & 2400 \\ 
J100200.96+591508.3   & 2015-12-04 &  0.55  &  148.5  & B & 1200 \\ 
J100551.19+125740.6   & 2015-12-04 &  0.55  &  60.1   & B & 1200 \\ 
J100935.66+265648.9   & 2015-12-05 &  0.59  &  45.5   & B & 1200 \\ 
J101440.21+192448.9   & 2015-03-25 &  0.58  & -57.58  & B & 1200 \\ 
J114343.76+550019.2   & 2015-03-25 &  0.58  & -117.58 & B & 1200 \\ 
J122342.82+581446.2*  & 2015-03-24 &  0.70  & -106.86 & C & 1200 \\ 
J130724.63+523715.2   & 2015-03-25 &  0.58  &  74.09  & B & 2400 \\ 
J131503.77+223522.7*  & 2015-03-24 &  0.70  &  -144.8 & C & 1200 \\ 
J134332.09+253157.7   & 2015-03-24 &  0.70  &  -21.33 & C & 2400 \\ 
J140116.03+542507.4   & 2015-03-25 &  0.58  &  81.76  & B & 1200 \\ 
J140228.72+091856.4   & 2017-06-24 &  1.17  &  120    & B & 1200 \\ 
J140510.39+114616.9   & 2017-06-25 &  0.85  &  12     & A & 1200 \\ 
J140735.47+503242.7   & 2017-06-25 &  0.85  &  163    & A & 2400 \\ 
J141208.47+102953.8   & 2017-06-24 &  1.17  &  0.0    & B & 1200 \\ 
J144252.78+205451.6   & 2017-06-24 &  1.17  &  101    & B & 1200 \\ 
J145843.39+113745.4   & 2017-06-25 &  0.85  &  19     & A & 1200 \\ 
J151116.53+233421.6   & 2017-06-25 &  0.85  &  165    & A & 1200 \\ 
J154603.78+031339.4   & 2017-06-24 &  1.17  &  31     & B & 1200 \\ 
J160839.57+120038.5   & 2017-06-25 &  0.85  &  174    & A & 1200 \\ 
J162307.88+391847.5   & 2017-06-24 &  1.17  &  80     & B & 1200 \\ 
J162335.06+454443.6   & 2017-06-25 &  0.85  &  40     & A & 1200 \\ 
J164428.48+435904.2   & 2017-06-24 &  1.17  &  82     & B & 1200 \\ 
J170639.14+334103.4   & 2017-06-25 &  0.85  &  156    & A & 1200 \\ 
J171759.66+332003.8   & 2017-06-24 &  1.17  &  68     & B & 1200 \\ 
J172125.92+281134.9   & 2017-06-24 &  1.17  &  168    & B & 1200 \\ 
J172208.82+280155.8   & 2017-06-24 &  1.17  &  130    & B & 1200 \\ 
J232028.21+150420.9*  & 2015-12-04 &  0.55  &  122.2  & B & 1200 \\ 
\hline
\end{tabular}
\caption{LRIS configuration for each object, using the 1$\arcsec$ slit placed along the semimajor axis of each galaxy.  For  the  blue  side  (LRIS-B),  we  used  the  600 groove mm$^{-1}$ grism blazed at 4000\AA, yielding a dispersion  of  0.63\AA\, pixel$^{-1}$.  We used three setups on the red side (LRIS-R):\newline
A: 600 groove mm$^{-1}$ grating blazed at 5000\AA, 5600\AA\, dichroic, yielding a dispersion of 0.80\AA\,pixel$^{-1}$\newline
B: 900 groove mm$^{-1}$ grating blazed at 5500\AA, 5600\AA\, dichroic, yielding a dispersion of 0.53\AA\,pixel$^{-1}$\newline
C: 1200 groove mm$^{-1}$ grating blazed at 7500\AA, 5000\AA\, dichroic, yielding a dispersion of 0.40\AA\,pixel$^{-1}$}
\label{tab:obs_setups}
\end{table*}
\newpage

The LRIS data were reduced using a Python pipeline to automate the standard IRAF reduction tasks.  Flexure on the red camera was corrected using the average shift in sky lines.  Sky lines are sparse in the wavelength range covered by the blue CCD, so each galaxy spectrum on the blue side was redshift-corrected using the redshift measured from the flexure-corrected red spectrum.  Flexure on the blue CCD was then calculated by comparing redshift-corrected galaxy emission lines with their expected rest frame values.  The longslit spectra were rectified along both the wavelength and spatial axes, yielding 2-dimensional spectra where each pixel row along the spatial axis is a fully reduced 1-dimensional spectrum.  

%
\section{Analysis}
\label{sec:analysis}

\subsection{Spatially Resolved Spectra}
\label{sec:binning}
\begin{center}
\begin{figure}
\includegraphics[width=\linewidth,clip]{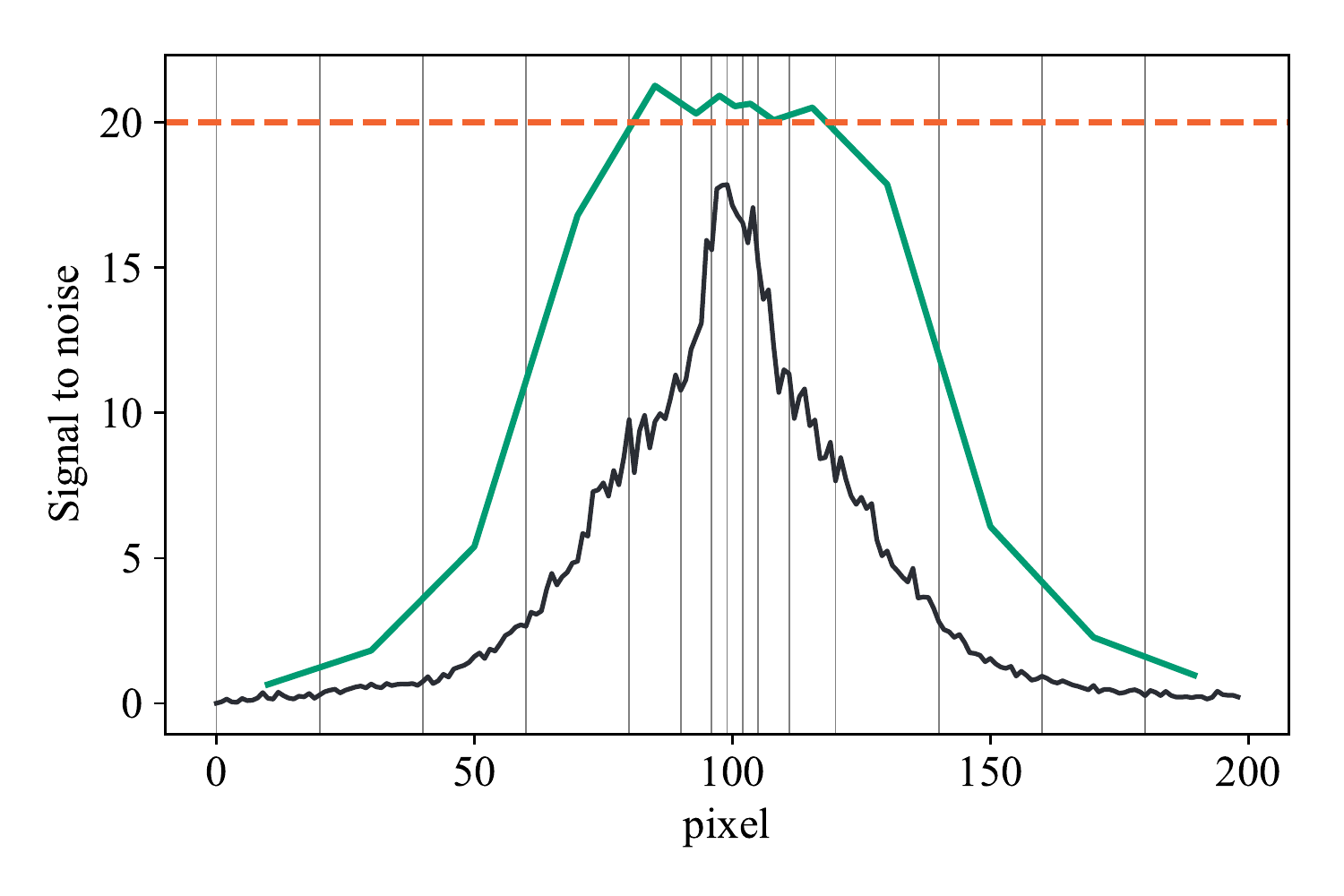}
\caption{Pixel number along the spatial axis is shown on the x axis.  The y axis shows the signal to noise ratio of a featureless portion of the spectrum red-ward of [\ion{O}{III}]$\lambda$5007.  Moving along the slit, we summed pixel rows (black) until the integrated spectrum's signal to noise ratio (green) reached the target S/N (red) or the maximum bin size was reached. The minimum bin size is 3 pixel rows, which sometimes results in even higher S/N.  Exposure times for each object were chosen with the intention of obtaining sufficient S/N in the outskirts, so the bin size is dependent on the observational setup used.
}
\label{fig:bin_example}
\end{figure}
\end{center}

%
The longslit spectra have been rectified along both the wavelength and spatial axes, creating a 2D spectrum where each pixel row is a fully reduced 1D spectrum.  Much of our analysis depends on fitting subtle spectral features, and thus require high signal to noise (S/N), especially in the dim outskirts of each galaxy.  To achieve the required S/N, we spatially bin the spectra, using larger bins on the outskirts of the galaxy, as shown in Figure \ref{fig:bin_example}.

Spectra were extracted along the slit by summing pixel rows until the target S/N ratio (typically 15$-$20) or maximum bin size was reached. In order to prevent summing the entire image and losing spatial data, the maximum bin size is set to be 20\% of the spatial axis.  The signal to noise was measured using the mean and standard deviation of a relatively featureless portion of the stellar spectrum, just red-ward of [\ion{O}{III}]$\lambda5007$.  The next bin would begin at the central pixel of the last bin.  Figure \ref{fig:bin_example} shows the signal to noise of each pixel row along the spatial axis (black). Vertical gray lines mark the divisions between bins, and the green line indicates the signal to noise when all pixel rows within each bin is summed.

\subsection{Emission Line Fluxes}
\label{sec:emission}
%
In order to measure accurate emission line fluxes, it is necessary to account for stellar absorption, which primarily affects the Balmer emission lines. The Penalized Pixel-Fitting software (\textsc{ppxf}) \citep{Cappellari2017} is used to fit and subtract the stellar continuum following the method described in \citet{Manzano2019} (henceforth Paper I), Section 3.1.  

After subtracting the best-fitting stellar population model of the galaxy, the residual emission lines were fit using a custom Bayesian MCMC maximum likelihood sampling algorithm, implemented using the Python package {\it emcee} \citep{emcee}.  A single-Gaussian model was used to fit emission lines for each galaxy in this sample, except in 13 cases when a second Gaussian component was needed (Paper I).  In these 13 cases the BPT flux ratios of the narrow component of the emission lines are shown in Figure \ref{fig:bpthist} and used when classifying these galaxies as AGN or star forming.


\subsection{Multi-Component Velocity Measurements}
\label{sec:ppxf}

In the \textsc{ppxf} software, emission lines are modeled as Gaussians and fit simultaneously with stellar templates.  Each stellar and gas template can be matched to a unique kinematic component, enabling the decomposition of multiple distinct line of sight velocities for stars and various species of emission lines.  Each spatially resolved velocity measurement presented in this work consists of three components: stellar, ionized hydrogen, and forbidden gas emission.

\begin{center}
\begin{figure}
\includegraphics[width=\linewidth,clip]{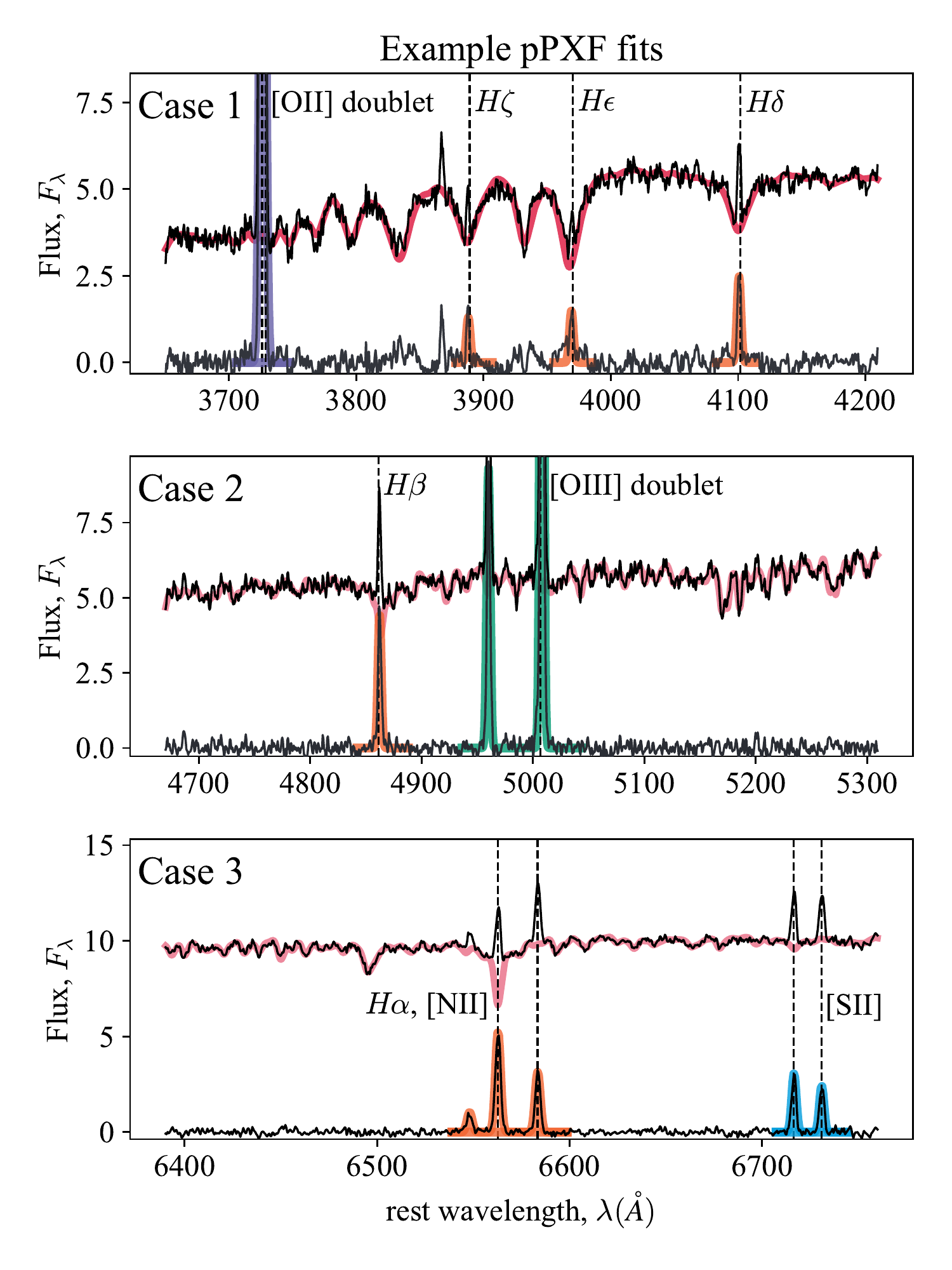}
\caption{Example \textsc{ppxf} fits to each of the three spectral sections used to determine velocity curves.  Each velocity curve consists of three components: stellar (red), hydrogen (orange), and forbidden (purple, green, and blue).}
\label{fig:fit_example}
\end{figure}
\end{center}

To avoid systematic errors introduced by wavelength calibration on the blue CCD, we fit $\sim 1000$\AA\, sections of the spectrum using \textsc{ppxf}.  Due to the strength of the H$\beta$ and [\ion{O}{III}] emission lines, and their proximity to the stellar feature MgIb, we measured line of sight velocities using the spectral region at rest wavelength $4500 - 5560$\AA\, whenever favorable.  Five galaxies were observed using the 5000\AA\, dichroic, which disrupts the H$\beta$, [\ion{O}{III}] region of the spectrum.  In these cases, stellar and gas kinematics were measured by fitting the portion of the spectrum containing [\ion{O}{II}] and  Balmer break ($3650 - 4550$\AA).  In cases where the [\ion{O}{III}] lines were faint, the fitting area containing H$\alpha$, [\ion{N}{II}], and [\ion{S}{II}] ($6400-6800$\AA) enabled gas velocity measurements to extend to a larger radius than fitting the $4500 - 5560$\AA\, region.  An example of fits to these three regions is shown in Figure \ref{fig:fit_example}.  

Thirteen galaxies in our sample have gas profiles that cannot be accurately modeled by a single Gaussian.  In these cases, two Gaussian components were used for each emission line: a narrow and a broad.  The velocity measurements presented in this work were obtained from the narrow Gaussian components.

\subsection{Circular Velocity Curves}
\label{sec:vcurves}
We extracted spatially resolved spectra along the semimajor axis of each galaxy by summing pixel rows, as described in Section \ref{sec:binning}.  Multicomponent line of sight velocity measurements were then taken from each spectrum following the method outlined in Section \ref{sec:ppxf}.  To convert line of sight velocity measured in \textsc{ppxf} into rotational velocity, we correct for the disc inclination angle (see Appendix \ref{app:incl}).  Zero velocity is measured at the galaxy's kinematic center, which is determined by the point of symmetry in the stellar velocity curve.  

\begin{figure}
	\includegraphics[width=\columnwidth]{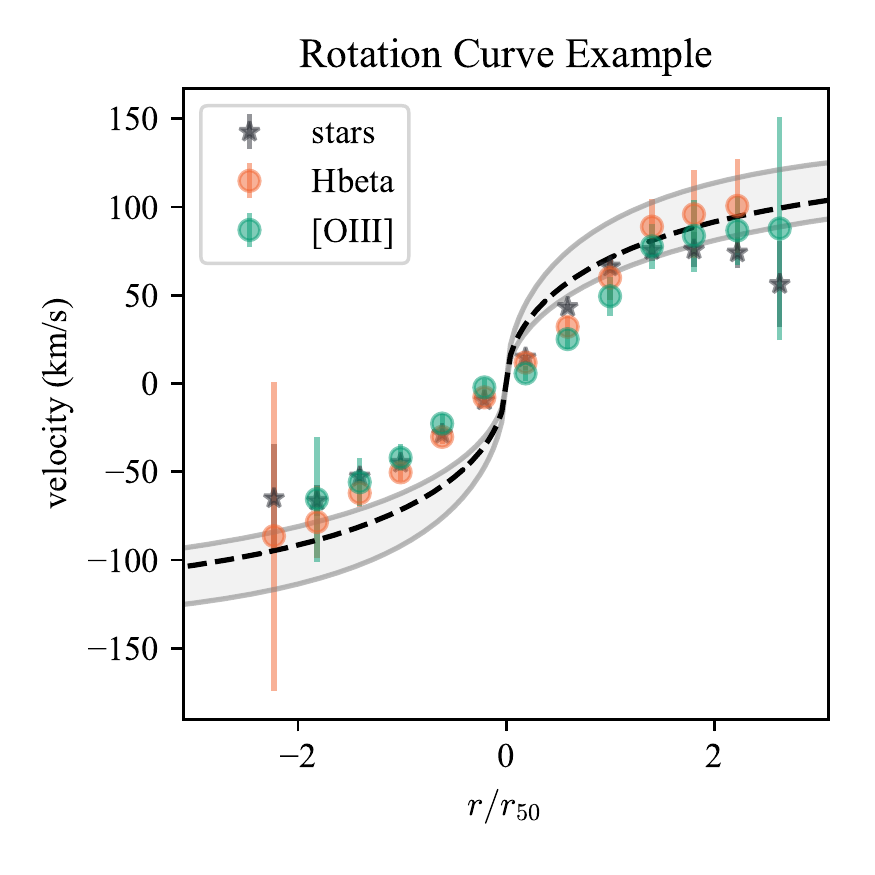}
    \caption{The object J170639.14+334103.4 was chosen randomly from our sample to provide an example of an orderly rotation curve, with a co-rotating disc of gas and stars.  The stellar (gray stars), H$\beta$ (orange circles) and [\ion{O}{III}] (teal circles) velocities are shown as a function of normalized radius, where $r_{50}$ is the r-band Petrosian 50\% radius.  The dotted black line indicates the expected velocity curve for an NFW profile with concentration parameter $c=10$.  The shaded gray region represents the NFW curve expected from halos with concentration parameter $c=8-15$.}
    \label{fig:rotation_example1}
\end{figure}

Figure \ref{fig:rotation_example1} shows the resulting rotational velocity curve for J170639.14+334103.4  as an example.  Stellar velocity measurements are plotted as gray stars and the ionized gas velocities for H$\beta$ and [\ion{O}{III}] are shown as orange and teal circles, respectively.  $r_{50}$ refers to the 50\% SDSS r-band Petrosian radius. The gray shaded region shows the analytic prediction of the rotation curve assuming the dark matter halo follows an NFW density profile (see Section \ref{subsec:nfw}).

%
\section{Results}
\label{sec:Results}
We obtained circular velocity curves for the stellar and ionized gas components of all 50 galaxies.  Five galaxies are excluded from the rest of this analysis based on the limited spatial extent of their stellar velocity curves.  For three galaxies (J024656.39-003304.8, J131503.77+223522.7, and J232028.21+150420.9), the spatial extent of their velocity curves were comparable (within 0.3\arcsec) with the seeing.  NGC 1569 was excluded because the slit only covered the central 0.25 kpc of the galaxy.  J122342.82+581446.2 was excluded because we were unable to achieve a sufficient S/N ratio to obtain more than two stellar velocity measurements. 
Rotation curves for the remaining 45 galaxies (26 AGN and 19 star forming) included in this analysis are shown in Appendix \ref{app:curves}.  


\subsection{Comparison with NFW}
\label{subsec:nfw}

\begin{figure}
	\includegraphics[width=\columnwidth]{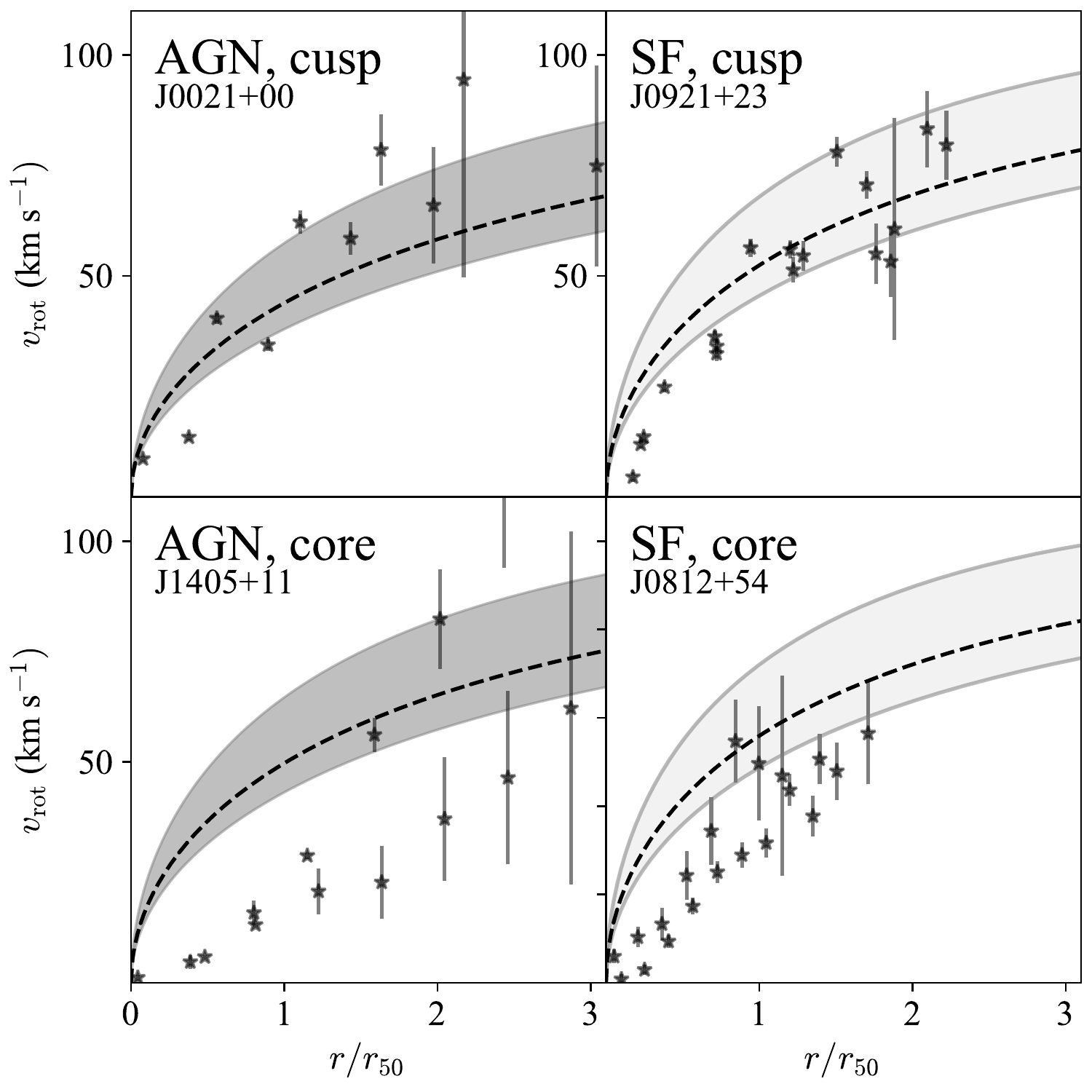}
    \caption{Four stellar absolute circular velocity curves are shown to demonstrate the variety of inner slopes found in our sample. Two examples of AGN are displayed in the left panels and two star forming galaxies are shown on the right.  Dashed black lines and shaded gray regions represent NFW velocity curves corresponding to a halo mass determined by abundance matching, with the MPA-JHU stellar mass as input.  Stellar velocity curves showing agreement with the expected NFW profile are shown on the top two panels. The velocity profiles in the bottom panels rise slowly, indicating cored dark matter density profiles.  }
    \label{fig:cusp-core}
\end{figure}

The $\Lambda$CDM model predicts that dark matter halos will follow an NFW profile, where the density steepens quickly in the inner regions and more slowly in the outer regions, forming a `cusp' profile.  However, observed dwarf galaxy rotation curves show a variety of slopes in their velocity profiles, some of which rise linearly with radius \citep{Oman2015}, implying underdense dark matter, or `cores,' in some galactic centres. This diversity in inferred dark matter profiles reported in a large number of observational studies and is yet to be explained by baryonic feedback models \citep[e.g.][]{Santos-Santos2020}.  
Since both AGN and stellar feedback have been shown to move large quantities of gas, and thus would be capable of altering the dark matter distribution of their host galaxies \citep[e.g.][]{Governato2010,Martizzi2013}, the process of forming cores might be a complex interplay between multiple feedback modes.  Any observed association between velocity curve shapes and AGN activity could help disentangle role of AGN in shaping dwarf galaxy dark matter halos. 

To aid in visual identification of cores in our sample,  we plot the NFW velocity curve of a $\Lambda$CDM halo corresponding to the predicted halo mass of each galaxy as gray shaded regions in Figures \ref{fig:rotation_example1}, \ref{fig:cusp-core}, and all other velocity curve plots in Appendix \ref{app:curves}.

It is difficult to observationally constrain the halo mass of a galaxy, so one popular approach is to use abundance matching \citep[e.g.][]{Moster2013}.  Abundance matching assumes a monotonic relation between stellar and halo mass and matches the cumulative abundance of galaxies on that relation.  We used abundance matching to estimate the halo mass corresponding to each galaxy's stellar mass reported in the MPA-JHU catalogue \citep{Brinchmann2004, kauffmann2003, tremonti2004}.  The only observational input is the stellar mass $M_\star$, so the halo mass $M_h$ can be estimated using abundance matching (\citet{Moster2013}, Equation 2):

\begin{equation}
M_* = M_h \left[2N\left(\frac{M_*}{M_1}\right)^{-\beta} + \left(\frac{M_h}{M_1}\right)^\gamma \right]^{-1}
\end{equation}
with four free parameters: N, the normalization parameter, a characteristic mass $M_1$ and low and high mass slopes $\beta$ and $\gamma$.

From these estimated halo masses, we estimate a virial radius 
\begin{equation}
r_v = \left(\frac{3}{4}\frac{M_h}{\pi v \rho_c^0}\right)^{1/3}
\end{equation} 

where $v = 200$ km/s, and $\rho_c^0 = 277.5 M_\odot/$kpc.  Assuming an NFW dark matter density profile, we construct a radial mass distribution (\citet{Lokas2001}, Equation 8):

\begin{equation}
M(r) = g(c) \left[ \ln(1+cs) - \frac{cs}{1+cs}   \right] M_h
\end{equation}

where $c$ is the concentration parameter, $s=r/r_v$, and $g(c) = [\ln(1+c)-c/(1+c)]^{-1}$.  We estimate the circular velocity curve from the radial mass distribution using $v(r) = \sqrt{G M(r)/r}$.  The dotted black line in Figure \ref{fig:rotation_example1} represents the rotation curve expected with an NFW profile with concentration parameter $c=10$, and the shaded gray region was calculated assuming a concentration parameter between $c=8,15$.  

There is no clear correlation between rotational velocity slopes and AGN activity in our sample.  Figure \ref{fig:cusp-core} shows examples of two AGN and two star forming stellar absolute velocity curves in varying levels of agreement with their predicted NFW profiles. The lack of any association between current AGN activity and central mass deficits could be attributed to a difference in time scales for AGN activity and core formation, or a number of other proposed factors not related to AGN \citep[e.g.][]{deBlok2010}.  A solution to the cusp-core dilemma appears to be beyond the scope of this work, and it remains to be seen whether AGN play a role in carving out cores in dwarf galaxies.


\subsection{Peculiar Gas}
\label{sec:peculiar}
\begin{figure}
	\includegraphics[width=\columnwidth]{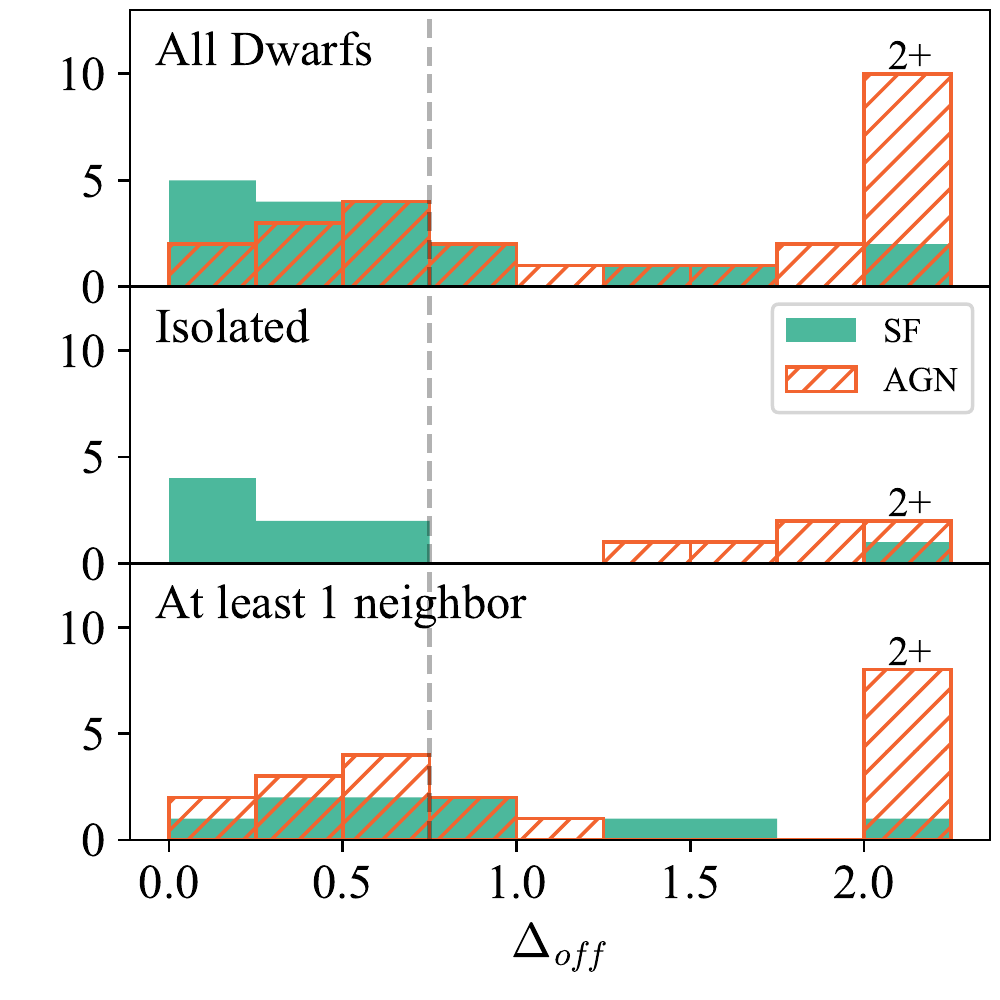}
    \caption{We define $\Delta_{\text{off}}$ to be the weighted absolute average of the velocity offset between stars and gas, divided by the average absolute stellar velocity.  This metric is used to quantify the degree of separation between the stellar and forbidden gas component.  Based on the bimodal distribution of  $\Delta_{\text{off}}$  in our sample, we consider galaxies $\Delta_{\text{off}} >= 0.75$ to have kinematically disturbed gas.}  
   \label{fig:del_off}
\end{figure}

\begin{figure*}
	\includegraphics[width=\textwidth]{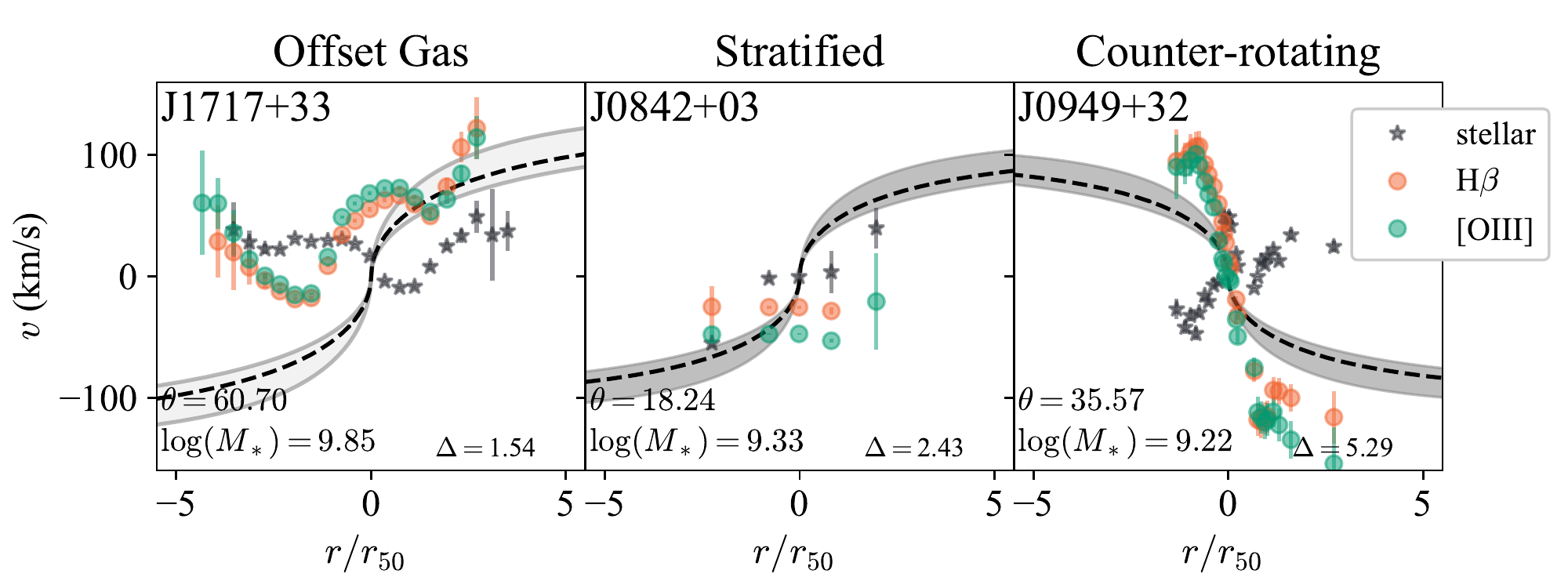}
    \caption{Examples of rotation curves where the gas is disturbed in different ways. 
    In some instances, gas is offset from the stars without showing any clear sign of rotation (left).  Some AGN have stratified narrow line components, where the Balmer and forbidden lines are kinematically distinct from one another (centre).  Five galaxies in our sample have gas and stellar discs rotating separately, and sometimes in opposite directions (right).}
    \label{fig:rotation_example}
\end{figure*}

The gas in these dwarfs show several distinct indications of non-rotational motion, which could be interpreted as inflows, outflows, or recently accreted gas.  To identify disturbed gas kinematics, we designed a metric to quantify the relative offset between the gas and stellar rotation curves.  $\Delta_{\text{off}}$ is the weighted average of the absolute velocity offset between the stellar and gas component, divided by the average absolute stellar velocity.

\begin{equation}
    \Delta_{\text{off}} = \dfrac{|\bar{\Delta}_{v}|}{|\bar{v}_*|}
\end{equation}

where the weighted average of the absolute velocity offset is 

\begin{equation}
   |\bar{\Delta}_{v}| = \dfrac{\sum_i |v_{*,i}-v_{\text{gas},i}|\,w_i}{\sum_i w_i}
\end{equation}

and the weights are the inverse of the combined errors of each velocity measurement, $w_i = (\Delta v_{*,i}^2 + \Delta v_{\text{gas},i}^2)^{-\frac{1}{2}}$

This calculation considers the offset between the stellar and forbidden gas component because the [\ion{O}{II}], [\ion{O}{III}], and [\ion{S}{II}] doublets are less affected by the subtraction of the stellar continuum and absorption than Balmer lines.  Faint broad lines associated with the AGN may also contribute to the line profiles, adding further uncertainty in velocities measured using Balmer emission lines. The forbidden [\ion{O}{II}], [\ion{O}{III}], and [\ion{S}{II}] doublets are not affected by such absorption or broad line region features and therefore give more accurate velocity measurements.

Histograms of $\Delta_{\text{off}}$ are shown in Figure \ref{fig:del_off} and individual $\Delta_\text{off}$ values are reported in the figures for each galaxy in Appendix \ref{app:curves}.  The top panel shows $\Delta_{\text{off}}$ for all 45 galaxies with rotation curves, the middle panel shows the distribution for all isolated galaxies, and the bottom shows the distribution for all galaxies that have at least one neighbour of comparable (or greater) mass (see Section \ref{sec:environment} for a discussion on galaxy environment).  The histograms show a bimodal distribution, which is most pronounced in the middle panel.  Based on this bimodality, we classify a galaxy as `disturbed' when $\Delta_{\text{off}} \geq 0.75$ (dotted gray line).  

Two galaxies (J010005.94-011059.0 and J090613.75+561015.5) have $\Delta_{\text{off}} < 0.75$ between the stars and forbidden emission lines, but the Balmer and forbidden gas components are offset from one another.  Following the same approach, we determine the threshold for offset Balmer and forbidden emission components to be (H$\beta$ - [\ion{O}{III}] $\Delta_{\text{off}} > 0.5$). Distinct kinematics associated with higher ionization emission lines is known as line stratification (see Section \ref{subsec:stratification}), and we count galaxies with stratified emission lines among the galaxies with disturbed gas.   By this criteria, 25 of the 45 galaxies with rotation curves are disturbed.   All rotation curves are plotted in Appendix \ref{app:curves}. The rotation curves showing orderly, co-rotating discs are shown in Figure \ref{fig:wellbehaved} and 25 disturbed rotation curves are shown in Figure \ref{fig:disturbed}.  

Of the 25 galaxies with disturbed gas, 19 host AGN and 6 do not. The majority (73\%) the AGN in our sample have disturbed gas, while only 32\% of star forming galaxies have $\Delta_\text{off} > 0.75$.  We noticed distinct properties in the non-rotational motion exhibited in the disturbed gas, and we show examples of rotation curves exhibiting different types of disturbances.  We observe gas that is generally offset from the stellar component, stratified emission lines, and counter-rotating gas. Examples of each type of disturbance are shown in Figure \ref{fig:rotation_example}.

\subsubsection{Line Stratification} 
\label{subsec:stratification}
In the stratified model of the narrow line region (NLR), lower ionization gas resides on the outer parts of the NLR while higher ionization lines are generated closer to the AGN \citep[e.g.][]{Croom2002,Andika2020}.  Distinct kinematics associated with emission lines of different ionization potentials imply a complex narrow line region that is  stratified in ionization and wind speed. We observe stratified gas components in seven AGN in our sample, with higher ionization lines showing higher velocities than lower ionization lines.  This implies a decelerating outflow: where a high velocity, high ionization wind is generated near the AGN, and while the low ionization gas in the outer region of the NLR flows more slowly.

We observe no correlation between wind speed and AGN luminosity, though the line stratification we observe in these seven galaxies preferentially occurs in galaxies that have outflows indicated by broad [\ion{O}{III}] components.  Plots of the stratified gas kinematics in these seven galaxies can be found in the Appendix, Figure \ref{fig:strat_curves}.

\subsubsection{Counter-Rotating}
\label{subsec:counter-rotating}
\begin{figure*}
	\includegraphics[width=\textwidth]{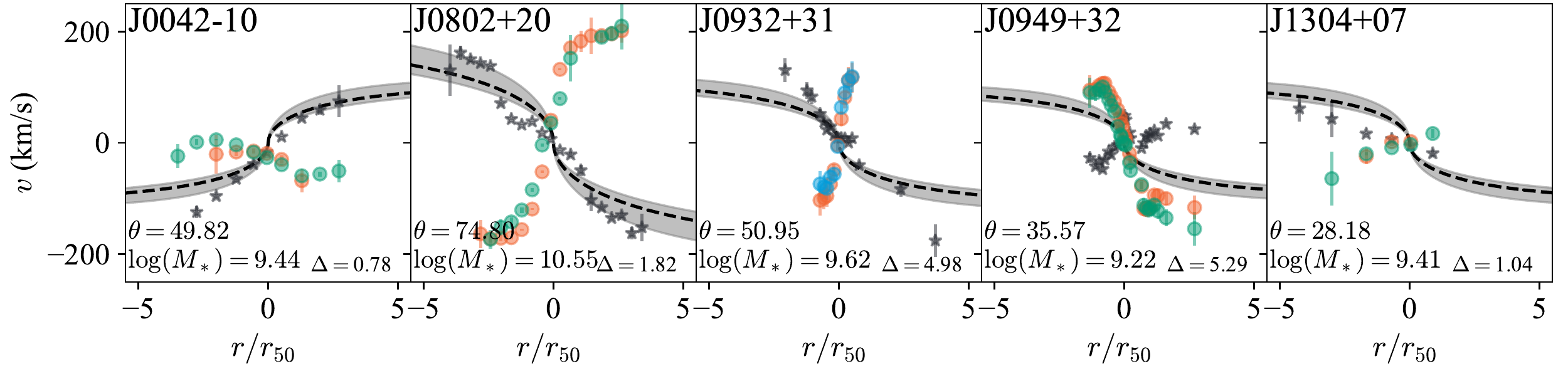}
    \caption{The rotation curves for five galaxies with counter-rotating gas are shown here.  The colours are as in Figure \ref{fig:rotation_example}, and the blue dots in the middle panel indicate gas velocities measured from the [SII] doublet.}
    \label{fig:cr_curves}
\end{figure*}

Of the 45 galaxies with rotation curves, five have counter-rotating gas and stellar components, shown in Figure \ref{fig:cr_curves}.  Counter rotating gas and stars have long been explained as the effect of the accretion of gas clouds or small satellites after the formation of the stellar disc \citep[e.g.][]{Thakar1996,Thomas2006,Katkov2014}.  As mentioned in the introduction, S19 found that counter-rotating gas in dwarf galaxies requires substantial gas removal, either via black hole feedback or environmental effects from fly-by interactions with more massive systems.  Interestingly, all five counter-rotating dwarfs in our sample host AGN.

S19 also make several predictions about the present-day properties of counter-rotating dwarfs.  In cases where re-accretion of gas is gradual, they find that counter-rotation can be very long lived (up to $\sim 2$ Gyr).  As a result, S19 predict no significant correlation with environment.  Three counter-rotating galaxies are isolated (J004214.99-104415.0, J080228.83+203050.2, J093251.11+314145.0) and two (J094941.20+321315.9, J130434.92+075505.0) have at least one neighbour with comparable mass.

An event that removes most of the original gas reservoir would quench star formation, making it likely that counter-rotating dwarfs will have older stellar populations and appear redder in colour.  All five counter-rotating galaxies discussed here have colours  $u-r > 2$, which is redder than the average for our sample ($u-r = 1.83$). S19 also find a 30\% gas deficit in counter-rotating dwarfs at $z=0$, relative to control galaxies at fixed stellar mass.  Though most of the galaxies in our sample lack gas mass estimates, we can turn our attention to the six galaxies with HI masses measured in the ALFALFA survey \citep{Bradford2018}.   Of these six, five are not disturbed and one is counter-rotating (J094941.20+321315.9).  The five co-rotating galaxies with HI measurements have a wide range of gas fractions ($\frac{M_{HI}}{M_*} \sim 0.23 - 7.3$), and the one counter-rotating has the smallest gas fraction ($\frac{M_{HI}}{M_*}* = 0.16$).  The gas component of J094941.20+321315.9 reaches velocities that exceed the predicted NFW curve, suggesting that this galaxy inhabits a more massive halo than predicted from its stellar mass.



\subsection{Star Formation Suppression}
\label{subsec:suppression}
\begin{figure}
	\includegraphics[width=\columnwidth]{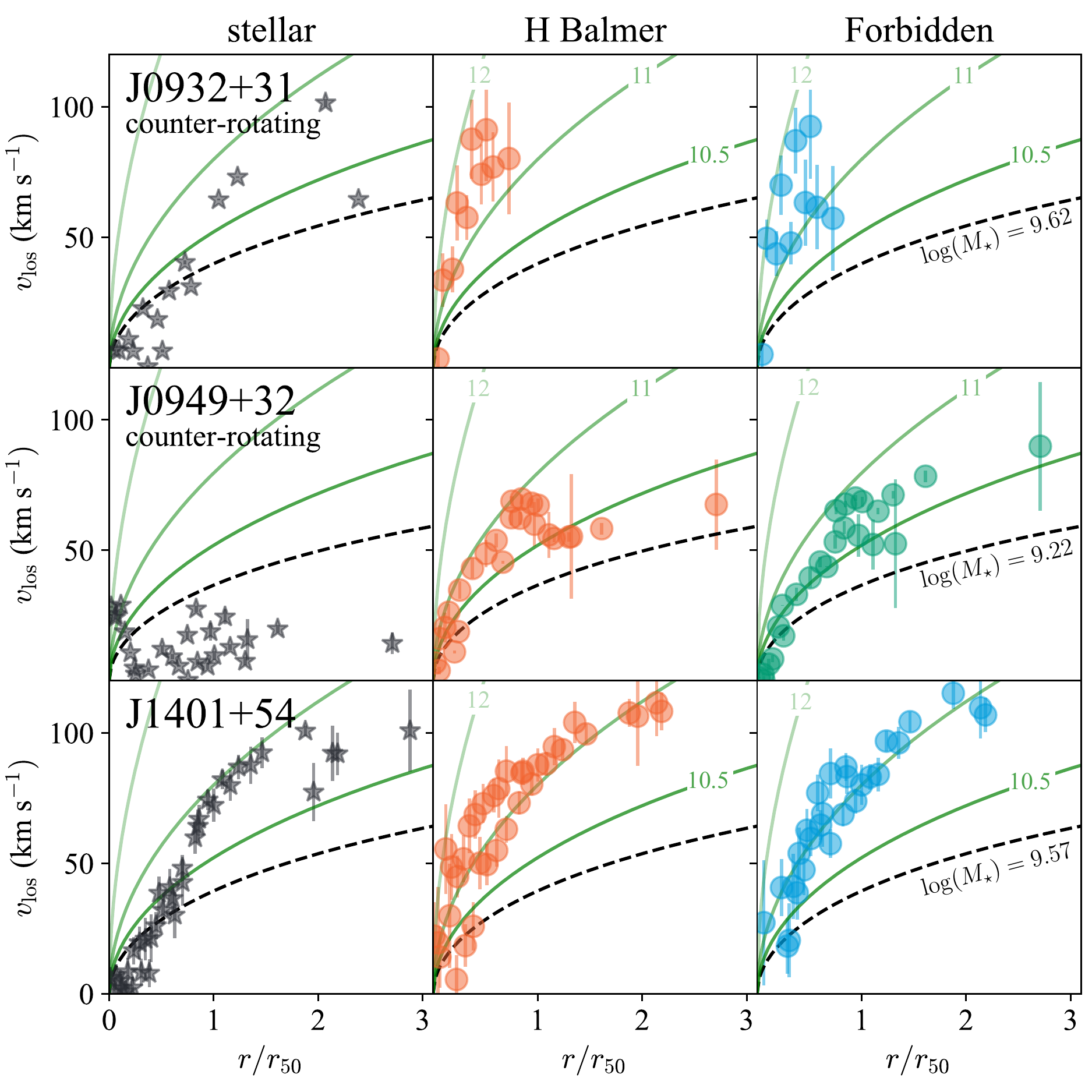}
    \caption{Absolute line of sight velocity $(v_\text{los})$ measurements that far exceed the expected NFW velocity curves based on their stellar masses indicate overmassive dark matter haloes relative to the measured stellar mass, implying ongoing star formation suppression.  Stellar (left), Balmer emission (centre) and forbidden emission (right) components are shown for the three galaxies in our sample where this is most apparent. Green curves mark expected NFW velocity curves for galaxies with stellar mass $\log(M_\star)=10.5, 11,\text{and } 12$ and concentration parameter $c=10$. Black dotted lines denote the expected NFW velocity curves based on each galaxy's MPA-JHU stellar mass.  All three of these galaxies host AGN, and two have counter-rotating gas components, which lends additional evidence potentially associating AGN with gas depletion and star formation suppression in dwarf galaxies.
    }  
   \label{fig:sf_supp}
\end{figure}


While abundance matching enables us to infer halo masses based on stellar masses, maximum line of sight velocity measurements can be used to place lower limits on the true halo mass of each galaxy.  If the measured lower limit of the halo mass of a galaxy exceeds the halo mass inferred from abundance matching to stellar mass, the galaxy can be said to have a lower star formation rate than expected.  Figure \ref{fig:sf_supp} shows absolute line of sight velocity curves for the three galaxies in our sample where this is most apparent.  As in previous figures, black dotted lines indicate the NFW curve predicted based on each galaxy's stellar mass. Green curves mark the NFW curves associated with stellar masses $\log(M_\star/M_\odot) = 10.5, 11, \text{and } 12$ and concentration parameter $c=10$.  The absolute line of sight velocity curves for the stellar (left), hydrogen Balmer emission (middle), and forbidden emission (right) components are shown. Line of sight velocities are plotted here in place of circular velocities to avoid errors introduced when correcting for disc inclination (Appendix \ref{app:incl}).  

The three galaxies shown in Figure \ref{fig:sf_supp} were selected from our sample using a similar approach to the one used to identify disturbed gas in Section \ref{sec:peculiar}.  We calculated the  offset between each kinematic component ($v_\text{los})$ and the predicted NFW curve $v_\text{NFW}$:
\begin{equation}
\Delta_{v_{\text{los}}\text{, NFW}} = \frac{\Sigma_i (v_{\text{los},i} - v_\text{NFW}) w_{\text{los},i}}{\Sigma_i w_{\text{los},i}}
\end{equation}
The distribution of $\Delta_{v_{\text{los}}\text{, NFW}}$ values for each component revealed three outlying galaxies that consistently fell above the threshold $\Delta_{v_{\text{los}}\text{, NFW}} > 0.5$: J093251.11+314145.0, J094941.20+321315.9, and J140116.03+542507.4.  

These galaxies have measured velocity curves that far exceed the expected NFW profile, indicating that they inhabit much larger halos than expected based on their stellar masses.  Their small stellar masses and red colours $(u-r \geq 2.23)$ suggest ongoing or recent star formation suppression.  It is interesting to note that all three galaxies host AGN, and two have counter-rotating gas, consistent with the scenario where AGN clear a substantial amount of gas from their hosts, limiting star formation.

\subsection{Environment}
\label{sec:environment}

To distinguish between the effects of environmental and secular processes on gas kinematics, we searched for luminous galaxies in the regions surrounding each dwarf.  Following the method and  criteria employed by \citet{Janz2017}, we queried the SDSS DR12 and 2MASS Redshift survey catalogues for luminous neighbours with $M_{K_S} < -21.5$ mag and SDSS $r < -16$ mag.  The completeness limit of this search is $M_{K_S} = -21.5$ mag at $z=0.02$; corresponding to $M_*\sim8\times10^9 M_\odot$ \citep{geha2012}. 
This limit is well below the magnitudes expected for galaxies in the mass range explored in this work, so extending the search to the redshift limit of our sample ($z=0.05$) has no effect on completeness.
The 2MASS Redshift Survey search revealed matches overlooked in the SDSS search due to missing redshifts, but SDSS optical counterparts existed for all 2MASS matches. We use SDSS DR12 g- and r-band photometry and the mass-magnitude relation of \citet{Bernardi2010} to estimate the masses of neighbouring galaxies. 

We identify galaxies within 1.5 Mpc and $\pm 1000\text{ km s}^{-1}$ of each dwarf in this sample as neighbours.  Using the g- and r-band mass-magnitude relation, we estimate the mass of each neighbour. In order to identify isolated galaxies, we count the number of neighbours with comparable mass (i.e. stellar mass $M_{\star,\text{neighbour}} \geq 0.75 \times M_{\odot,\text{dwarf}}$) within 1.5 Mpc and $\pm 1000\,\text{km}\,\text{s}^{-1}$.  The number of neighbours for each dwarf can be found in Table \ref{tab:sample}.  

A recent study of 62,258 dwarf ($M_\star < 5\times10^9 M_\odot$) galaxies finds no discernible difference between the environments of AGN and non-AGN dwarf galaxies, suggesting that environmental factors may not play a dominant role in triggering AGN \citep{Kristensen2020}.  Similarly, we find no statistically significant link between AGN and environment.  However, \citet{Kristensen2020} suggest that remnants of past interactions may be reflected in gas kinematics.

The importance of environment in disturbing gas kinematics can be explored by considering the bottom panel of Figure \ref{fig:del_off}, which shows the distribution of $\Delta_{\text{off}}$ for galaxies with at least one neighbour of comparable (or greater) mass within 1.5 Mpc.  Star forming and AGN-hosting galaxies with neighbours show varying degrees of gas disturbance, further supporting the notion that environment may not play a dominant role in triggering AGN activity.  The most extreme values of $\Delta_{\text{off}}$ are associated with AGN, suggesting that AGN are capable of generating large scale disturbances in their host galaxies' ISM, though the presence of AGN-hosting galaxies with undisturbed gas implies this is not always the case.  

On the other hand, isolated galaxies show a distinct bimodal distribution, where star forming galaxies tend to have orderly rotating discs, and disturbed gas is almost always associated with AGN.  Table \ref{tab:isolated} lists the seven disturbed, isolated galaxies, six of which host AGN.  Two of these six isolated AGN have counter rotating gas, and three have outflows indicated by distinct broadened components in [\ion{O}{III}]$\lambda$5007, as in Paper I.  To clarify, the outflows in these four galaxies are identified with multicomponent Gaussian fits to the [\ion{O}{III}] line profile; disturbed gas is determined here by a velocity offset between the narrow component of the emission lines from the stellar component.

The single isolated and disturbed star forming galaxy, J101440.21+192448.9, was originally included in our sample based on broad lines in H$\alpha$ \citep{Reines2013}, which later faded, likely due to transient stellar activity \citep{baldassare2016}.  This object is classified as an outflow galaxy in Paper I based on broadened wings in its [\ion{O}{III}] emission line profile.  Follow-up Integral Field Unit (IFU) observations with KCWI revealed that the rotation axis of J101440.21+192448.9 is in fact parallel to the slit position in our LRIS observations (Liu et al., in prep).  Since the slit is not aligned with the stellar disc, these velocity measurements do not reflect the true rotational velocity of this galaxy.  

Follow-up KCWI observations were obtained for eight objects in this sample (Liu et al., in prep). These IFU observations produce line of sight velocities that agree with our longslit measurements and confirm that in all other cases, the LRIS slit was oriented perpendicular to the rotational axes, as intended.  Proper slit orientation for the remainder of the sample should be verified with similar follow-up IFU observations, though the confirmation of correct slit placement in all but one galaxy is encouraging. Since J101440.21+192448.9 is an active star forming galaxy with supernova-driven outflows, it is reasonable to surmise that bright star forming regions could outshine the rest of the disc, confusing the photometric fit used to determine slit placement. This calls into question the slit placement for other galaxies with non-rotational kinematics and signs of active star formation, such as J171759.66+332003.8.

\begin{table*}
\centering
\begin{tabular}{| l c c c c c c c c c c c c|}
\hline
Abbreviated & Redshift  & $\log(M_*)$  & $\log(M_{\text{BH}})$& $r_{50}$  & BPT & HeII & neigh- & dist-   & counter- & outflow & strat- \\ 
Name        &           &              &                      & (kpc)     &     &      & bors   & urbed   & rotating &         & ified \\ 
(1)         & (2)       & (3)          & (4)                  & (5)       & (6) & (7)  & (8)    & (9)     & (10)     & (11) & (12) \\ \hline
J0021+00    & 0.0180 &  9.15        & ---                  & 0.89      &  AGN       & AGN    &   7  & -- & -- & -- & -- \\
J0042-10    & 0.0359 &  9.44        & ---                  & 1.34      &  AGN       & AGN    &   5  & \m & \m & -- & -- \\
J0100-01    & 0.0515 &  9.44        & ---                  & 0.99      &  Comp.     & Comp.  &   5  & \m & -- & \m & -- \\
J0156-00    & 0.0450 &  9.39        & ---                  & 1.91      &  SF        & SF     &  13  & \m & -- & -- & -- \\
J0246-00    & 0.0464 &  9.35        & 5.7$^c$              & 1.34      &  AGN       & AGN    &  37  & -- & -- & -- & -- \\
J0300+00    & 0.0095 &  8.72        & ---                  & 2.11      &  SF        & SF     &   0  & -- & -- & -- & -- \\
NGC 1569    & $-80\text{ km s}^{-1}$ &  8.56 & ---         & 7.09      &  SF        & SF     &  --  & -- & -- & \m & -- \\
J0755+24    & 0.0290 &  8.85        & ---                  & 1.78      &  SF        & SF     &   3  & -- & -- & -- & -- \\
J0802+10    & 0.0145 &  9.60        & 3.6$^b$              & 1.29      &  AGN       & AGN    &   3  & -- & -- & -- & -- \\
J0802+20    & 0.0286 &  10.6        & ---                  & 1.97      &  Comp.     & AGN    &   0  & \m & \m & -- & -- \\
J0811+23    & 0.0157 &  9.06        & 4.4$^c$              & 0.58      &  AGN       & AGN    &   2  & \m & -- & \m & \m \\
J0812+54    & 0.0086 &  9.29        & ---                  & 1.49      &  SF        & SF     &   0  & -- & -- & -- & -- \\
J0840+18    & 0.0149 &  9.13        & 4.3$^c$              & 0.94      &  AGN       & AGN    &   2  & \m & -- & \m & -- \\
J0842+03    & 0.0289 &  9.33        & ---                  & 0.75      &  Comp.     & AGN    &   2  & \m & -- & \m & \m \\
J0851+39    & 0.0407 &  9.28        & 5.4$^c$              & 1.70      &  AGN       & AGN    &   2  & -- & -- & -- & -- \\ 
J0906+56    & 0.0465 &  9.37        & 5.4$^c$              & 1.51      &  AGN       & AGN    &   5  & \m & -- & \m & -- \\
J0911+61    & 0.0263 &  8.95        & ---                  & 2.03      &  SF        & SF     &   7  & \m & -- & -- & -- \\      
J0921+23    & 0.0281 &  9.27        & ---                  & 1.27      &  SF        & SF     &   0  & -- & -- & -- & -- \\
J0932+31    & 0.0153 &  9.62        & 3.5$^b$              & 0.58      &  AGN       & AGN    &   0  & \m & \m & -- & -- \\          
J0948+09    & 0.0103 &  8.73        & 4.2$^b$              & 0.73      &  Comp.     & AGN    &   5  & \m & -- & -- & \m \\
J0949+32    & 0.0051 &  9.22        & 4.1$^b$              & 0.58      &  Comp.     & AGN    &   6  & \m & \m & -- & -- \\
J0954+47    & 0.0326 &  9.47        & 4.9$^c$              & 1.75      &  AGN       & AGN    &   0  & \m & -- & \m & -- \\
J1002+59    & 0.0093 &  9.60        & 4.1$^b$              & 1.12      &  AGN       & Comp.  &   1  & \m & -- & -- & \m \\
J1005+12    & 0.0093 &  9.64        & 4.8$^b$              & 0.97      &  AGN       & AGN    &   0  & \m & -- & \m & -- \\
J1009+26    & 0.0143 &  8.75        & 5.1$^c$              & 0.62      &  AGN       & AGN    &   0  & \m & -- & \m & -- \\
J1014+19    & 0.0284 &  8.56        & ---                  & 0.88      &  SF        & SF     &   0  & \m & -- & \m & -- \\           
J1143+55    & 0.0269 &  8.92        & ---                  & 1.03      &  SF        & SF     &   1  & -- & -- & -- & -- \\         
J1223+58    & 0.0146 &  9.39        & 6.1$^b$              & 1.01      &  AGN       & AGN    &   6  & -- & -- & -- & -- \\         
J1304+07    & 0.0479 &  9.41        & ---                  & 1.13      &  AGN       & AGN    &  16  & \m & \m & -- & -- \\  
J1307+52    & 0.0259 &  9.10        & ---                  & 1.18      &  SF        & SF     &   2  & -- & -- & -- & -- \\      
J1315+22    & 0.0226 &  9.20        & ---                  & 1.58      &  SF        & SF     &   1  & -- & -- & -- & -- \\        
J1343+25    & 0.0280 &  9.19        & ---                  & 3.09      &  SF        & SF     &   3  & -- & -- & -- & -- \\        
J1401+54    & 0.0059 &  9.57        & 3.2$^b$              & 0.59      &  AGN       & AGN    &   5  & -- & -- & -- & -- \\      
J1402+09    & 0.0195 &  8.83        & ---                  & 0.79      &  AGN       & AGN    &  26  & \m & -- & -- & -- \\     
J1405+11    & 0.0178 &  9.23        & ---                  & 1.15      &  AGN       & AGN    &   4  & -- & -- & -- & -- \\
J1407+50    & 0.0070 &  8.72        & ---                  & 0.77      &  Comp.     & AGN    &   8  & -- & -- & -- & -- \\ 
J1412+10    & 0.0324 &  8.99        & ---                  & 0.56      &  AGN       & AGN    &   3  & \m & -- & -- & -- \\
J1442+20    & 0.0426 &  8.89        & ---                  & 0.89      &  AGN       & AGN    &   1  & \m & -- & \m & \m \\
J1458+11    & 0.0198 &  9.91        & ---                  & 2.40      &  SF        & SF     &   0  & -- & -- & -- & -- \\
J1511+23    & 0.0143 &  9.68        & ---                  & 1.01      &  SF        & SF     &   0  & -- & -- & -- & -- \\
J1546+03    & 0.0132 &  9.49        & ---                  & 2.05      &  SF        & AGN    &   3  & -- & -- & -- & -- \\
J1608+12    & 0.0166 &  9.75        & ---                  & 2.65      &  SF        & SF     &   0  & -- & -- & -- & -- \\
J1623+39    & 0.0172 &  9.22        & ---                  & 2.68      &  SF        & SF     &   2  & \m & -- & -- & -- \\
J1623+45    & 0.0064 &  9.41        & ---                  & 0.82      &  AGN       & AGN    &   0  & \m & -- & -- & -- \\
J1644+43    & 0.0178 &  9.53        & ---                  & 0.86      &  SF        & SF     &   0  & -- & -- & -- & -- \\
J1706+33    & 0.0301 &  9.41        & ---                  & 1.95      &  SF        & SF     &   7  & -- & -- & -- & -- \\
J1717+33    & 0.0151 &  9.85        & ---                  & 1.03      &  SF        & SF     &   1  & \m & -- & \m & -- \\
J1721+28    & 0.0280 &  10.0        & ---                  & 3.49      &  SF        & SF     &   0  & -- & -- & -- & -- \\
J1722+28    & 0.0281 &  9.23        & ---                  & 1.71      &  SF        & SF     &   3  & \m & -- & -- & -- \\
J2320+15    & 0.0130 &  9.57        & 3.7$^b$              & 0.59      &  AGN       & AGN    &   6  & -- & -- & -- & -- \\
\hline
\end{tabular}
\caption{
(1) * indicates objects excluded from this study.
(2) Redshift is calculated by fits to stellar absorption lines in LRIS spectra using \textsc{ppxf}.
(3) Stellar mass reported in the MPA-JHU catalogue.
(4) $^b$ Black hole mass lower limits derived using Eddington Luminosity arguments (M14) $^c$ Black hole mass estimated using as $M_{BH}\propto R\Delta V^2/G$, where $\Delta V$ is measured from broad H$\alpha$, extrapolating the BLR radius -- luminosity relation extends into the low mass regime (R13)
(5) SDSS Petrosian radius containing 50\% of r-band flux, in kiloparsecs.
(6) Classification of dominant ionization source based on Gaussian fits to emission lines using the process described in Section \ref{sec:emission}.  The spectra were extracted from the central 1kpc of each galaxy.
(7) Same as (6) but for HeII classification (S15)
(8) Number of galaxies with comparable mass within 1.5 Mpc. 
(9) Classified as disturbed (see Section \ref{sec:peculiar})
(10) Gas is clearly rotating and offset from stellar component. 
(11) Has an outflow identified by an additional broad component in as defined in Paper I.
(12) Narrow emission lines are stratified (see Section \ref{subsec:stratification})}
\label{tab:sample}
\end{table*}

\begin{center}
\begin{table}
\begin{tabular}{|c|c|c|c|c|c|}
    name &  $\log(M_\star)$ & AGN & counter-  & outflow & u$-$r \\ 
    &&& rotating & (Paper I) &colour  \\\hline
J0802+20   & 10.6 & \m & \m &    &  2.39 \\
J0932+31   &  9.6 & \m & \m &    &  2.23 \\
J0954+47   &  9.5 & \m &    & \m &  2.20 \\
J1005+12   &  9.6 & \m &    & \m &  2.16 \\
J1009+26   &  8.7 & \m &    & \m &  1.91 \\
J1014+19   &  8.6 &    &    & \m &  1.25 \\ 
J1623+45   &  9.4 & \m &    &    &  2.13 \\\hline

\end{tabular}
\caption{Seven galaxies in our sample have disturbed gas kinematics and no neighbours of comparable mass within 1.5 Mpc and $\pm 1000$ km s$^{-1}$.  Six of them host AGN, two have counter-rotating gas, and four have outflows indicated by broad wings in their [OIII]5007 emission line profiles.  $u-r$ colour is based on SDSS cModelMag photometry.}
\label{tab:isolated}
\end{table}
\end{center}


%
\section{Summary}
\label{sec:Summary}
From Keck LRIS longslit spectroscopy, we measured rotational velocity curves of 45 dwarf galaxies.  Our sample consists of  26 galaxies with AGN and a control sample of 19 star forming galaxies with no optical or IR evidence of AGN.  The rotation curves are decomposed into stellar, Balmer emission, and forbidden emission components.  In order to investigate the potential effects of AGN on gas kinematics in this sample, we quantified velocity offsets between stellar and gas components.  We summarize our conclusions below.


\begin{enumerate}


\item We detect counter-rotating gas in five of 45 galaxies, and AGN are present in all five cases. A study of the occurrence, properties, and evolutionary history of counter-rotating galaxies in Illustris \citep{Starkenburg2019} finds that removal and re-accretion of gas is necessary for counter-rotating gas discs to form, and that periods of AGN activity are sometimes associated with large drops in gas mass.  We find that the properties of the counter-rotating dwarfs in our sample agree with present-day properties of counter-rotating dwarf galaxies predicted from Illustris.

\item We use the weighted average of velocity offsets between stellar spectra and emission lines to identify disturbed gas kinematics. We find disturbed gas in 25 out of 45 galaxies in our sample.  Of the 26 AGN in our sample, 19 (or 73\%) have disturbed gas.  Star forming galaxies tend to have orderly, co-rotating gas discs, with only 32\% showing disturbed gas.  

\item At least three galaxies in our sample have line of sight velocities far exceeding those expected based on their stellar masses.  These line of sight velocities constitute a lower limit estimate of true halo mass, indicating that these galaxies inhabit much more massive dark matter haloes than expected based on their small stellar masses. This potential evidence of ongoing star formation suppression is most apparent in three AGN-hosting galaxies within our sample, two of which have counter-rotating gas.  This suggests that AGN could be associated with gas removal and star formation suppression in dwarf galaxies.  A detailed study of the star formation histories of these galaxies is warranted.

\item In the absence of environmental influence, kinematically disturbed gas is expected to be caused by secular processes.  Fifteen dwarfs in our sample are isolated, with no neighbouring galaxies of comparable mass within 1.5 Mpc and $\pm1000$ km s$^{-1}$.  Of these, seven galaxies have disturbed gas.  We find that six out of seven isolated galaxies with disturbed gas host AGN. 

\end{enumerate}

Our findings imply that AGN play an important, and perhaps dominant, role in disturbing gas and limiting star formation in dwarf galaxies.  This represents additional evidence of the importance of AGN-driven winds in dwarf galaxy evolution and further highlights the importance of including AGN feedback in galaxy formation models.  A detailed follow-up IFU study of ionized gas kinematics in a subset of these galaxies is in progress (Liu et. al., in prep). Multiphase gas kinematic measurements, gas mass measurements, and comparison with simulations run with detailed AGN feedback models are necessary for a full understanding of AGN feedback in dwarf galaxies.

\vspace{1cm}
{\bf Acknowledgements}
We would like to thank the anonymous referee,
whose careful reading and thoughtful feedback 
has helped to improve and clarify this manuscript. 
Support for this program was provided
by the National Science Foundation, under
grant number AST 1817233.  Additional support
was provided by NASA through a grant from the
Space Telescope Science Institute (Program AR-
14582.001-A), which is operated by the Association of
Universities for Research in Astronomy, Incorporated,
under NASA contract NAS5-26555.  
The data presented herein were obtained at the W. M.
Keck Observatory, which is operated as a scientific partnership 
among the California Institute of Technology,
the University of California and the National Aeronautics 
and Space Administration. The Observatory was made 
possible by the generous financial support of the
W. M. Keck Foundation.
The authors wish to recognize and acknowledge the
very significant cultural role and reverence that the 
summit of Mauna Kea has always had within the Indigenous
Hawaiian community. We are most fortunate to have the
opportunity to conduct observations from this mountain.
Some of the data presented herein were obtained using the UCI
Remote Observing Facility, made possible by a generous gift 
from John and Ruth Ann Evans. 
This research has made use of the NASA/IPAC Extragalactic Database (NED), which is operated by the Jet Propulsion Laboratory, California Institute of Technology, under contract with the National Aeronautics and Space Administration.

{\bf Data Availability}\\
The data underlying this article will be shared on reasonable request to the corresponding author.
\bibliographystyle{mnras}
\bibliography{master}

\appendix

\section{Inclination Correction}
\label{app:incl}
\begin{figure}
	\includegraphics[width=\columnwidth]{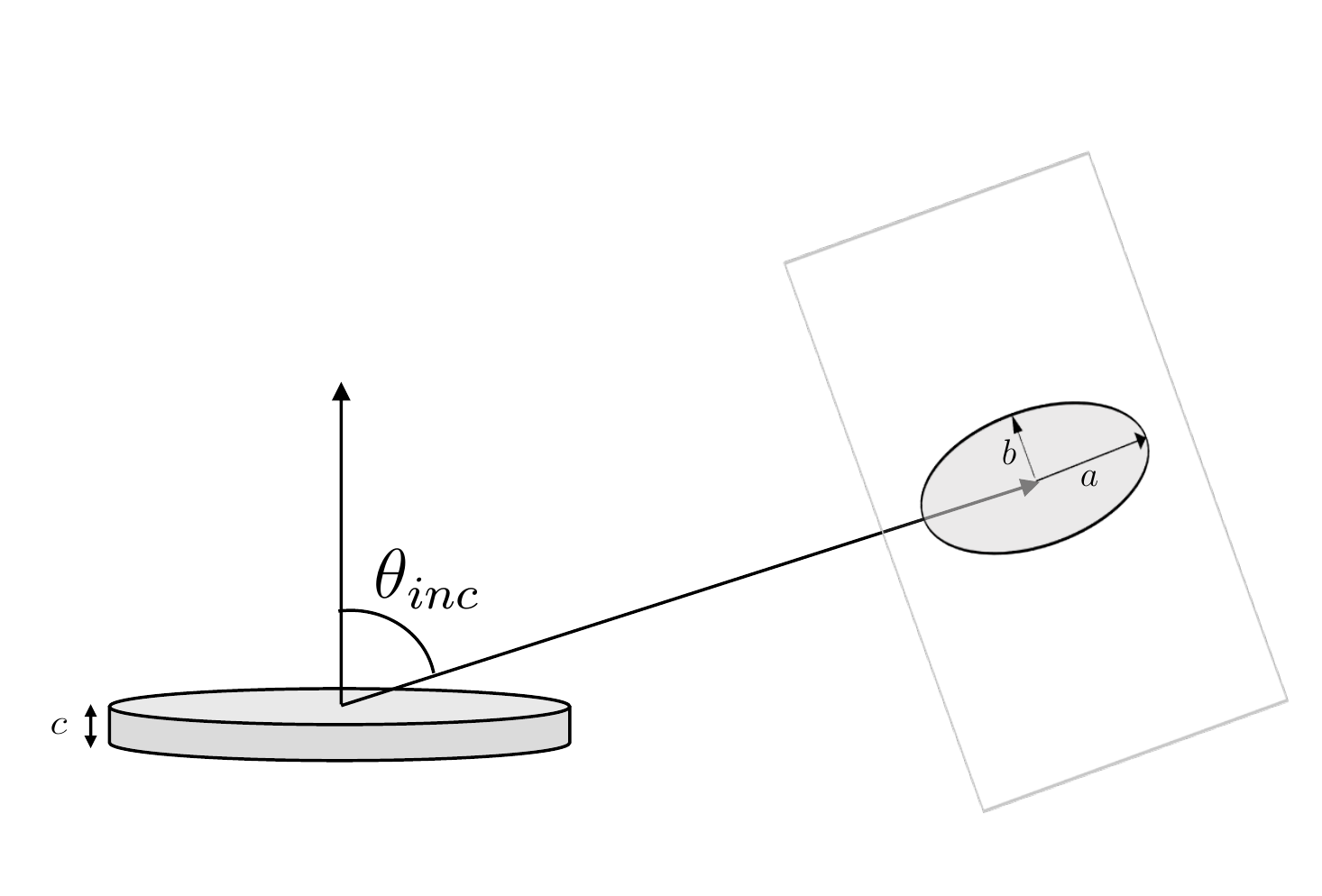}
    \caption{A diagram to illustrate how a disc with thickness $c$ appears as an ellipse when viewed from the inclination angle $\theta_\text{inc}$.  The inclination angle can be estimated by measuring the ellipse semimajor and minor axes $a$, $b$ and assuming the disc thickness $c$.}
    \label{fig:incl}
\end{figure}

\textsc{ppxf} returns line of sight velocities.  In order to translate the line of sight velocities measured in \textsc{ppxf} into rotational velocities, the inclination angle must be accounted for.  The inclination angle $\theta_{\text{inc}}$ of a galaxy is defined as the angle between the line of sight and the normal to the disc.  A perfectly round disc, inclined by $\theta_{\text{inc}}$ is projected as an ellipse in an image, with semimajor axis $a$ and semiminor axis $b$.  Assuming an infinitely thin disc, $\theta_{\text{inc}} = \cos^{-1}(b/a)$. More realistically, we can introduce a thickness parameter $q = c/a$, which describes a circular disc of radius $a$ and thickness $c$, as illustrated in Figure \ref{fig:incl}.  The projection of this shape when imaged is still approximately elliptical, and the inclination angle can more accurately be estimated using 
\begin{equation}
\cos^{-1}(\theta_\text{inc}) = \frac{(b/a)^2-q^2}{1-q^2}
\end{equation}

The ellipticity $(b/a)$ was obtained by a fit to the SDSS $r-$band image using the Python package photutils \citep{Bradley2019}.  In the absence of any constraint on galaxy morphology, we used a range of $q = 0.1 - 0.2$
\citep{Giovanelli1994}.

It is straightforward to translate line of sight velocity to circular velocity using the inclination angle: 
\begin{equation}
v_\text{rot} = \frac{v_\text{los}}{\sin(\theta_\text{inc})}
\end{equation}

The range of inclination angles then translates to a range of possible $v_\text{rot}$ values.  The $v_\text{los}$ measurements are corrected using the mean value for $\theta_\text{inc}$.  The minimum and maximum values for $\theta_\text{inc}$ are used to estimate the error introduced to the $v_\text{rot}$ measurements by the uncertainty in inclination.

We forego inclination correction for two cases (J081145.29+232825.7 and J084234.51+031930.7) where the inclination angle is less than 20 degrees.  These galaxies are viewed face-on, and show no distinct rotation.  Their gas kinematics imply a stratified narrow line region with a non-rotational, possibly spherical outflow (see Section \ref{subsec:stratification}).   If this is the case, line of sight velocities are more physically meaningful than rotational velocities.

\section{All rotation curves}
\label{app:curves}

In this appendix, we present all 45 rotation curves included in this analysis.  Each curve was measured following the process described in Section \ref{sec:analysis}.  Spatially resolved spectra were extracted as described in Section \ref{sec:binning}, and kinematic measurements were obtained by spectral fits performed using \textsc{ppxf} (Section \ref{sec:ppxf}).  Each rotation curve consists of three components: stellar velocity (gray stars), gas velocity from Balmer emission lines (orange circles), and gas velocity from forbidden emission lines.  Whenever possible, the forbidden emission line velocities are measured from [\ion{O}{III}] (teal circles).  [\ion{O}{II}] (purple circles) were used for spectra obtained using the 5000\AA\, dichroic.  When gas emission was faint (H$\beta$ equivalent width less than 0.5\AA\, or [\ion{O}{III}] equivalent width less than 2.0 \AA), we use longer wavelengths to perform kinematic measurements.  In this case, Balmer gas velocities were measured from H$\alpha$ and [\ion{S}{II}] (blue circles) were used to measure the forbidden emission line component.  

Gray shaded regions denote the rotation curve predicted by assuming an NFW dark matter profile, with halo mass obtained using abundance matching, with the MPA stellar mass as input (see Section \ref{sec:Results}). The spatial axis is normalized by the SDSS r-band petrosian radius, $r_{50}$.  

The velocity measurements and associated errors are corrected for disc inclination as described in Appendix \ref{app:incl}.  The angle relative to the normal of the disc and the stellar mass, as listed in the MPA-JHU catalogue, are shown in the bottom left of each curve plot.  $\Delta_{\text{off}}$ is shown on the bottom right.

Asymmetric drift, or the difference between the local circular speed and the mean rotation of the galaxy, acts to reduce the observed rotational velocity due to the presence of randomly oriented and non-circular orbits.  In the absence of gas surface density measurements, we do not correct for asymmetric drift.

\begin{figure*}
	\includegraphics[width=\textwidth]{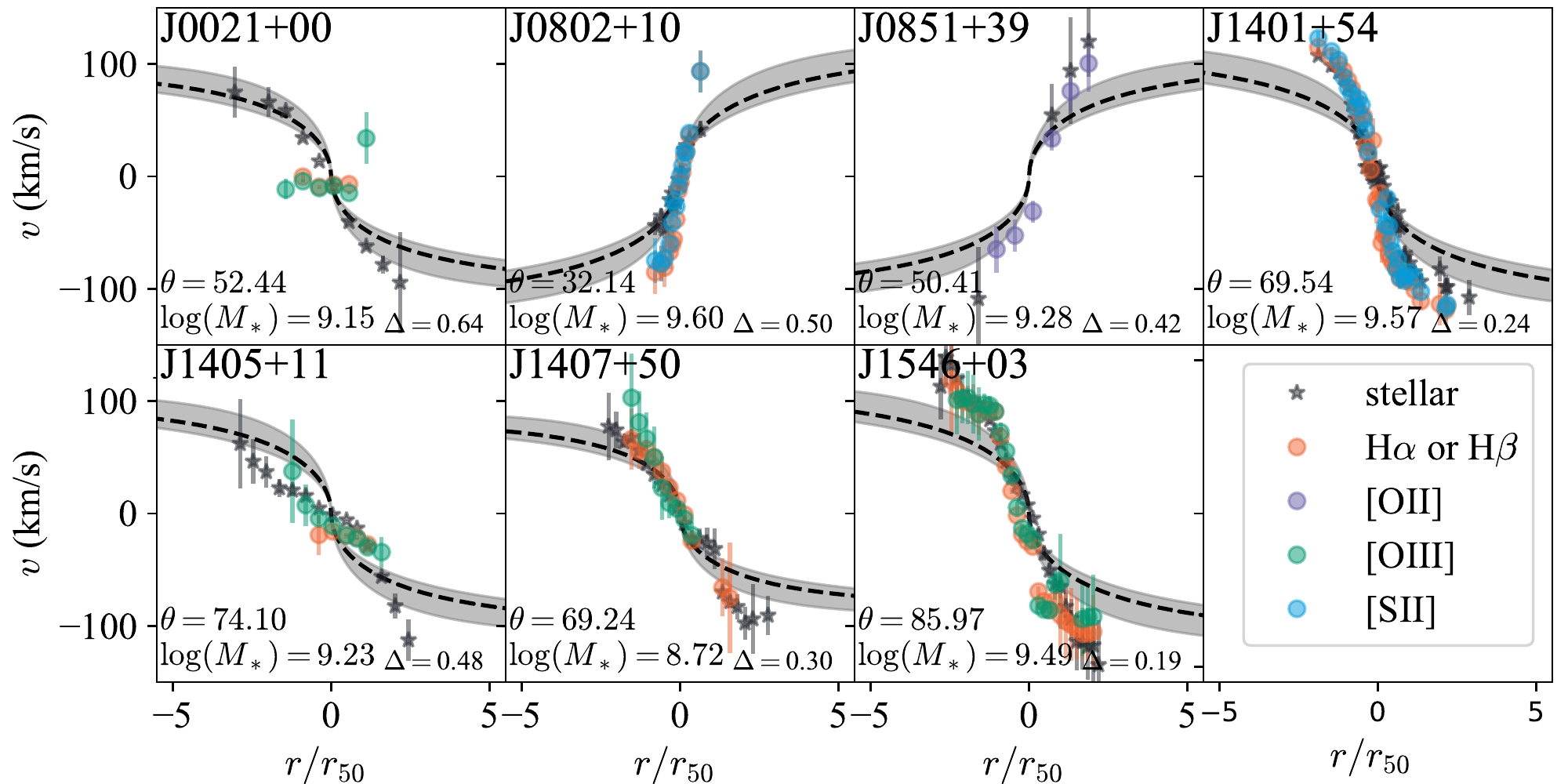}
	\includegraphics[width=\textwidth]{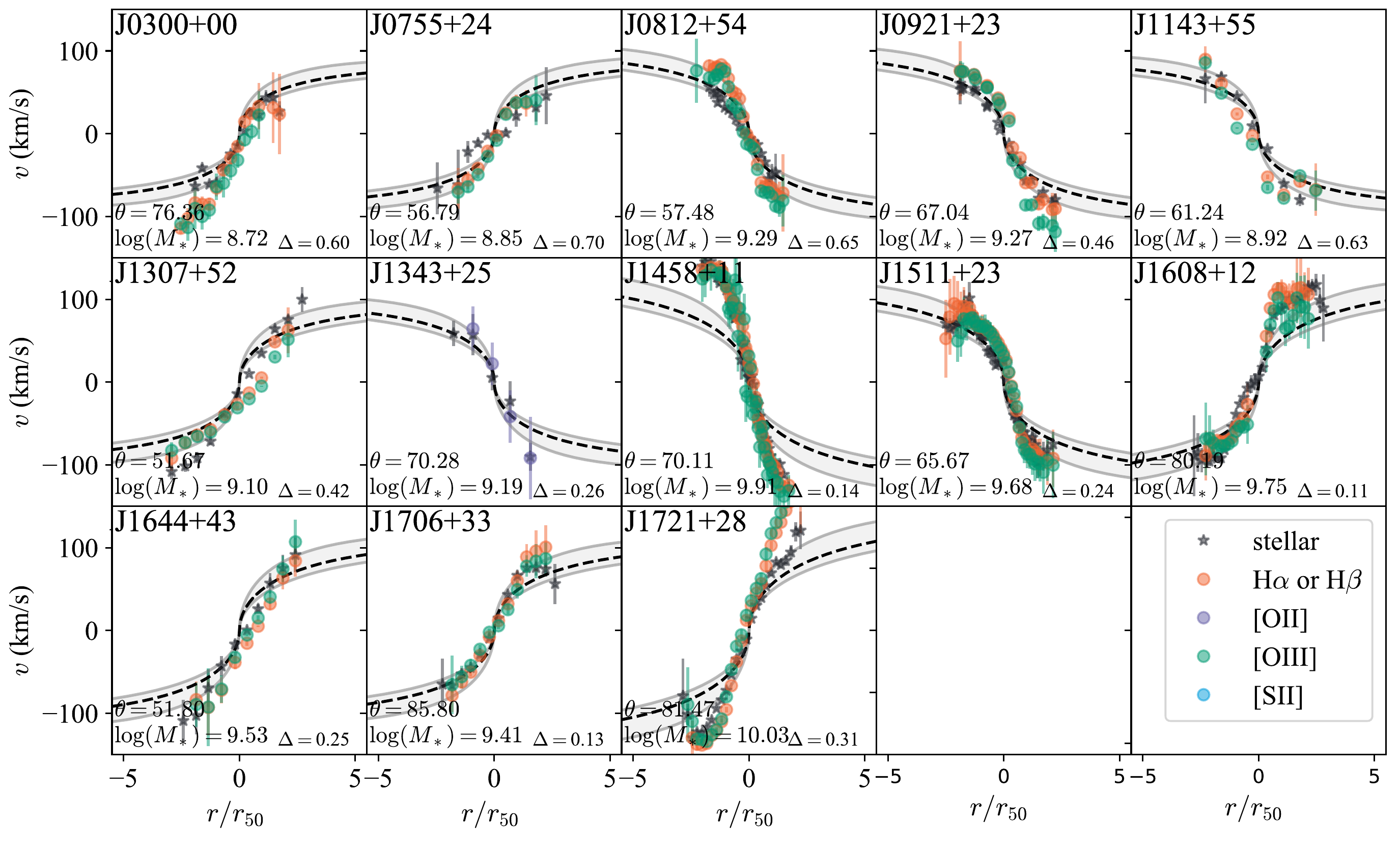}
    \caption{Twenty of the 45 galaxies with rotation curves have gas that rotates with the stellar component. The shaded regions denote velocity curves following NFW profiles, assuming concentration parameters between $c=8,15$.  Galaxies hosting AGN are shown in the top panel and star forming galaxies are in the bottom.}  
   \label{fig:wellbehaved}
\end{figure*}

\begin{figure*}
	\includegraphics[width=\textwidth]{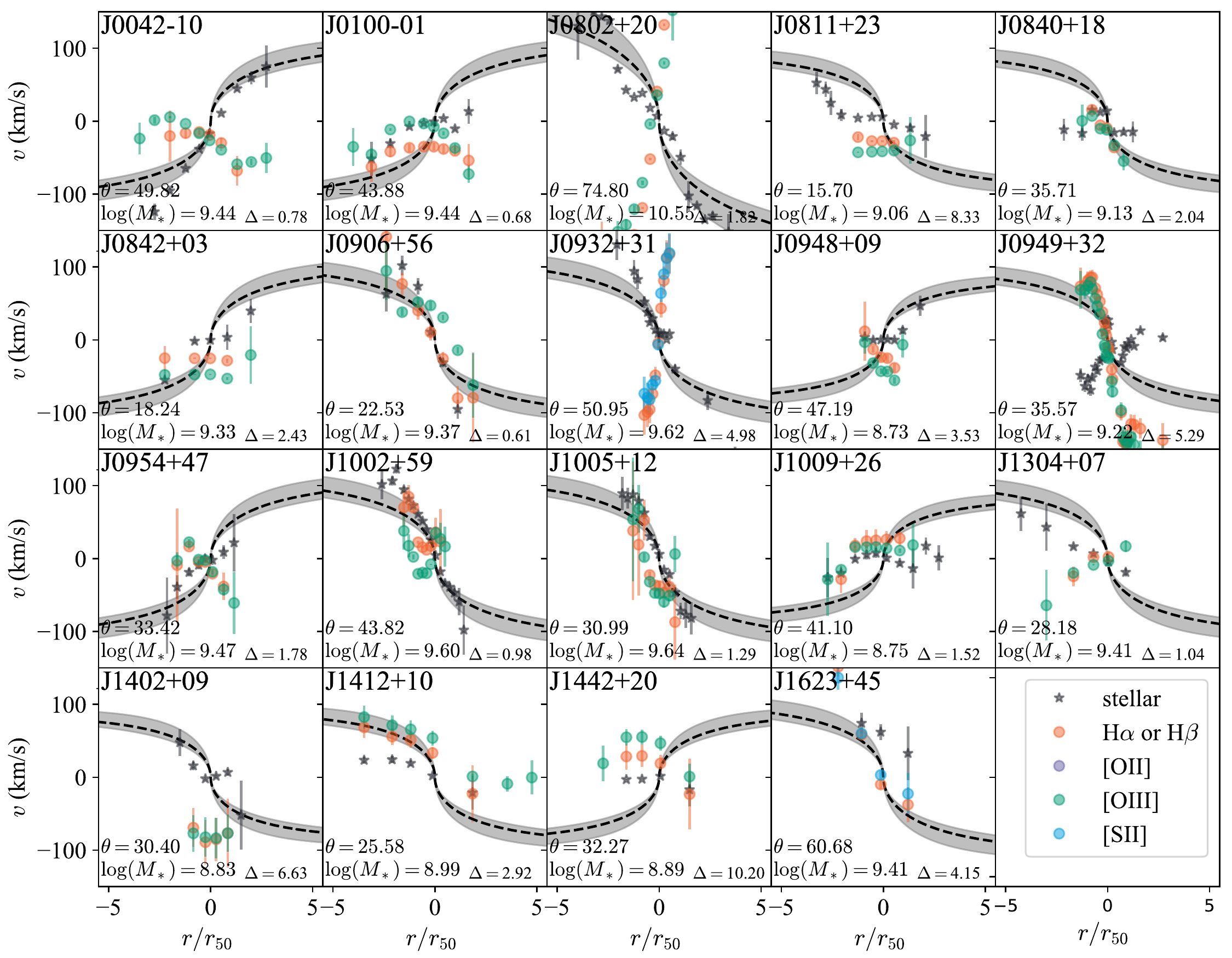}
	\includegraphics[width=\textwidth]{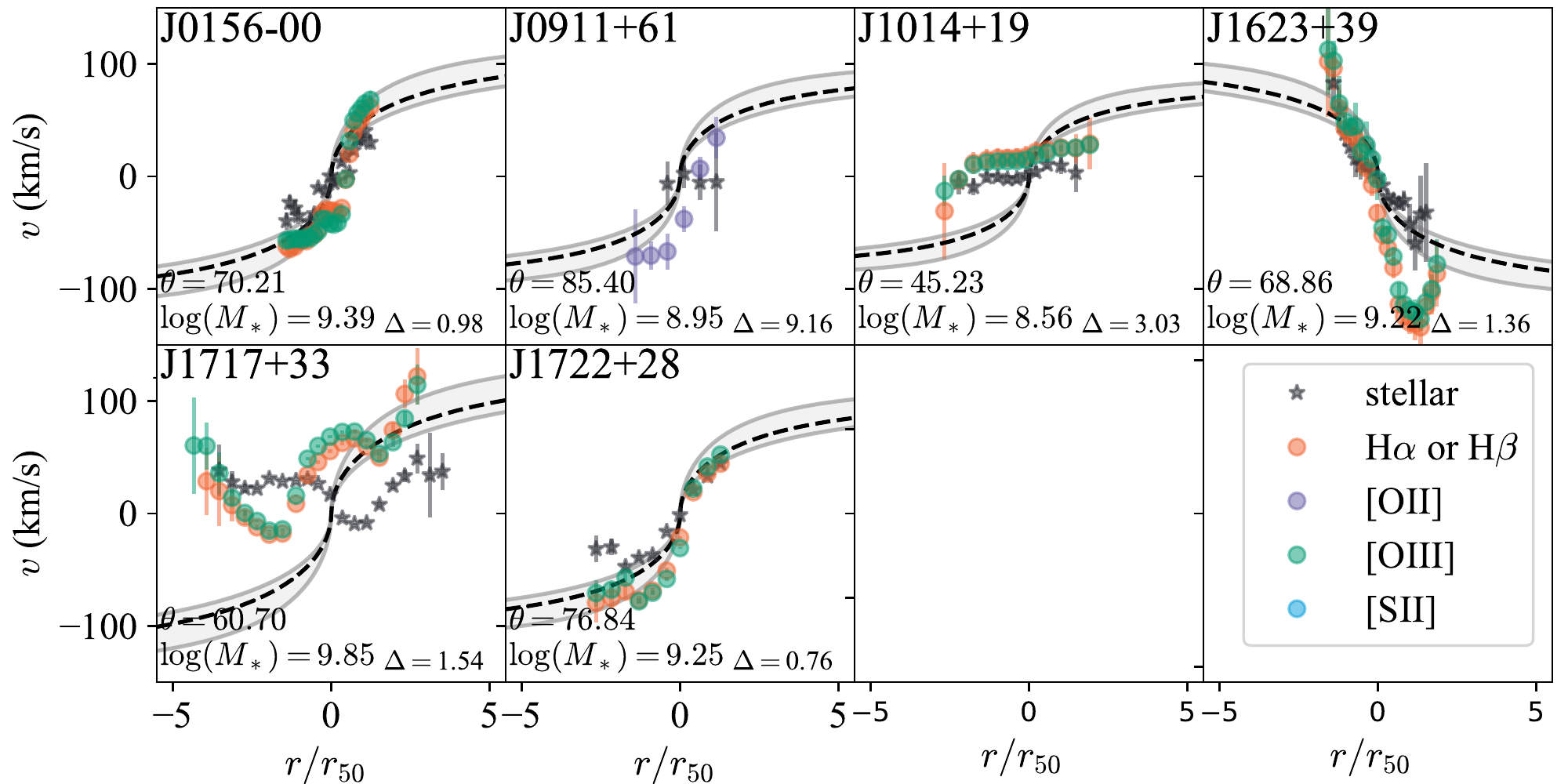}
    \caption{Twenty-five of the 45 galaxies with rotation curves have disturbed gas.  See Section \ref{sec:peculiar} for a discussion on how we classify galaxies as disturbed.  Galaxies hosting AGN are shown in the top group and star forming galaxies are in the bottom.}  
   \label{fig:disturbed}
\end{figure*}

\begin{figure*}
	\includegraphics[width=\textwidth]{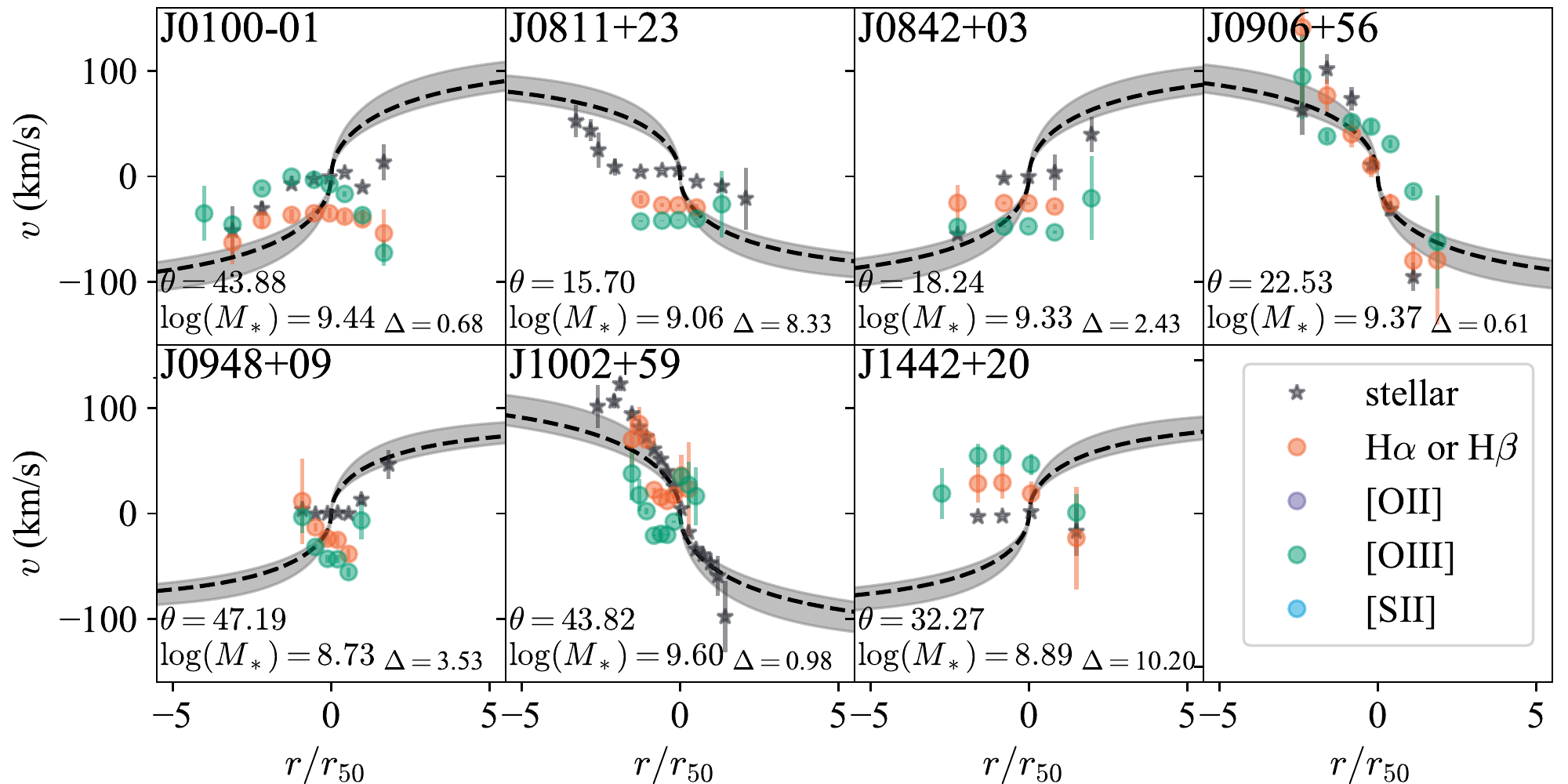}
    \caption{Eight of the 45 galaxies with rotation curves show stratification in their emission lines.  See Section \ref{sec:peculiar} for a discussion on how we identify line stratification. Line of sight velocity gradients are shown for J0811+23 and J0842+03, as they are considered face-on based on their small inclination angles ($\theta < 20 ^{\circ}$) and thus are not corrected for inclination (as described in Appendix \ref{app:curves}).  Rotational velocity curves (corrected for disc inclination) are shown for all other galaxies in this figure.}
    \label{fig:strat_curves}
\end{figure*}

\end{document}